\documentclass[3p]{elsarticle}
\journal{Radiology: Artificial Intelligence}

\usepackage[utf8]{inputenc} % allow utf-8 input
\usepackage[T1]{fontenc}    % use 8-bit T1 fonts
\usepackage{textcomp}
\usepackage{helvet}
\usepackage{url}            % simple URL typesetting
\usepackage{booktabs}       % professional-quality tables
\usepackage{amsfonts}       % blackboard math symbols
\usepackage{amsmath}        % 
\usepackage{amssymb}        % 
\usepackage{amsthm}         % 
\usepackage{nicefrac}       % compact symbols for 1/2, etc.
\usepackage{microtype}      % microtypography
\usepackage{xcolor}         % colors
\usepackage{lipsum}  
\usepackage{graphicx}
\usepackage{epsfig}
\usepackage{tikz}
\usetikzlibrary{external}
\usepackage{pgfplots}
\pgfplotsset{compat=newest}
\usepackage[labelfont=bf]{caption}
\usepackage[caption=false]{subfig}
\usepackage{float}
\usepackage{lineno}
\begin{document}

\begin{frontmatter}

\title{Deep Learning Segmentation of Ascites on Abdominal CT Scans \\
for Automatic Volume Quantification}

\author[1]{Benjamin Hou}
\author[2]{Sung-Won Lee}
\author[3]{Jung-Min Lee}
\author[4]{Christopher Koh}
\author[5]{\\Jing Xiao}
\author[6]{Perry J. Pickhardt}
\author[1]{Ronald M. Summers\corref{corrauth}}

\address[1]{Radiology and Imaging Sciences, Clinical Center -}
\address[3]{Women’s Malignancies Branch, National Cancer Institute -}
\address[4]{Liver Diseases Branch, National Institute of Diabetes and Digestive and Kidney Diseases -\\- National Institutes of Health, Bethesda, MD, USA.}
\address[5]{Ping An Technology, Shenzhen, China.}
\address[2]{The Catholic University of Korea, Seoul St. Mary's Hospital}
\address[6]{Department of Radiology, University of Wisconsin, School of Medicine \& Public Health, Madison, WI, USA.}

\cortext[corrauth]{Corresponding author (email: \url{rms@nih.gov})}

\begin{abstract}

\noindent\textbf{Purpose:} To evaluate the performance of an automated deep learning method in detecting ascites and subsequently quantifying its volume in patients with liver cirrhosis and ovarian cancer. 

\noindent\textbf{Materials and Methods:} This retrospective study included contrast-enhanced and non-contrast abdominal-pelvic CT scans of patients with cirrhotic ascites and patients with ovarian cancer from two institutions, National Institutes of Health (NIH) and University of Wisconsin (UofW). The model, trained on The Cancer Genome Atlas Ovarian Cancer dataset (mean age, 60 years $\pm$ 11 [s.d.]; 143 female), was tested on two internal (NIH-LC and NIH-OV) and one external dataset (UofW-LC). Its performance was measured by the Dice coefficient, standard deviations, and 95\% confidence intervals, focusing on ascites volume in the peritoneal cavity.

\noindent\textbf{Results:} On NIH-LC (25 patients; mean age, 59 years $\pm$ 14 [s.d.]; 14 male) and NIH-OV (166 patients; mean age, 65 years $\pm$ 9 [s.d.]; all female), the model achieved Dice scores of 0.855$\pm$0.061 (CI: 0.831-0.878) and 0.826$\pm$0.153 (CI: 0.764-0.887), with median volume estimation errors of 19.6\% (IQR: 13.2-29.0) and 5.3\% (IQR: 2.4-9.7) respectively. On UofW-LC (124 patients; mean age, 46 years $\pm$ 12 [s.d.]; 73 female), the model had a Dice score of 0.830$\pm$0.107 (CI: 0.798-0.863) and median volume estimation error of 9.7\% (IQR: 4.5-15.1). The model showed strong agreement with expert assessments, with $r^2$ values of 0.79, 0.98, and 0.97 across the test sets.

\noindent\textbf{Conclusion:} The proposed deep learning method performed well in segmenting and quantifying the volume of ascites in concordance with expert radiologist assessments. 

\end{abstract}

% \begin{keyword} %alphabetical order
% Abdomen \sep
% Ascites \sep
% CT \sep
% Convolutional Neural Networks \sep
% Deep Learning Algorithms \sep
% Segmentation \sep
% Supervised Learning \sep
% Volumetric Analysis
% % \MSC[2020] Primary 62H12 \sep
% % Secondary 62F12
% \end{keyword}

\end{frontmatter}

\vspace{\baselineskip}
\noindent This manuscript has been accepted for publication in Radiology: Artificial Intelligence \\ (\url{https://pubs.rsna.org/doi/10.1148/ryai.230601}), which is published by the \\ Radiological Society of North America ({\textsf{\copyright}} 2024 RSNA).

\newpage

\noindent\textbf{Summary Statement:}

\noindent The deep learning segmentation model accurately delineated and estimated volume of ascites regions in patients with liver cirrhosis and ovarian cancer, demonstrating strong concordance with expert manual measurements. 

\vspace{\baselineskip}

\noindent\textbf{Key Points:}

\noindent 1. The deep learning segmentation model achieved high accuracy in ascites volumetric quantification across both internal (n=25, F1 0.855$\pm$0.061 [CI: 0.831, 0.878]; n=166, F1 0.826$\pm$0.153 [CI: 0.764, 0.887]) and external (n=124, F1 0.830$\pm$0.107 [CI: 0.798, 0.863]) datasets. 

\noindent 2. The model demonstrated strong concordance with expert radiologists ($r^2$ values of 0.79, 0.98, and 0.97 across the three test datasets). 

\vspace{\baselineskip}

\noindent\textbf{Abbreviations:}

\noindent $-\hspace{0.1em}$ DL: deep learning \\
\noindent $-\hspace{0.1em}$ LC: liver cirrhosis \\
\noindent $-\hspace{0.1em}$ OV: ovarian cancer \\
\noindent $-\hspace{0.1em}$ SD: standard deviation \\
\noindent $-\hspace{0.1em}$ CI: confidence interval 

\vspace{\baselineskip}

\noindent\textbf{Author Contributions:}

\noindent Guarantor of integrity of entire study, B.H., S.W.L.; study concepts/study design or data acquisition or data analysis/interpretation, all authors; manuscript drafting or manuscript revision for important intellectual content, all authors; approval of final version of submitted manuscript, all authors; agrees to ensure any questions related to the work are appropriately resolved, all authors; literature research, B.H., J.M.L., C.K., J.X.; clinical studies, J.M.L., C.K., P.J.P.; experimental studies, B.H., R.M.S.; statistical analysis, B.H.; and manuscript editing, B.H., S.W.L., J.M.L., C.K., J.X., P.J.P.

\vspace{\baselineskip}

\noindent\textbf{Acknowledgments:}

\noindent This research was supported by the Intramural Research Program of the National Institutes of Health, Clinical Center, and utilized the computational resources of the National Institutes of Health high-performance computing Biowulf cluster. Huge thanks to Dr Pritam Mukherjee for his invaluable support and insightful guidance during the writing of this manuscript. The results shown here are in whole or part based upon data generated by the TCGA Research Network: \url{https://cancergenome.nih.gov/}

\vspace{\baselineskip}

\noindent\textbf{Disclosures of conflicts of interest:} 

\noindent B.H. No relevant relationships. S.W.L. No relevant relationships. J.M.L. No relevant relationships. C.K. No relevant relationships. J.X. No relevant relationships. P.J.P. Consulting fees, Bracco Diagnostics, Nanox-AI, GE HealthCare; stock or stock options, SHINE, Elucent, and ColoWatch. R.M.S. Grant, PingAn; royalties or licenses, iCAD, Philips, ScanMed, PingAn, Translation Holdings, MGB; support for travel to an Advisory Group meeting, Duke University; Radiology: Artificial Intelligence editorial board member.

\newpage

\section{Introduction}

Ascites is a medical condition characterized by the abnormal accumulation of fluid in the abdominal cavity, specifically within the peritoneal space. It is commonly associated with various underlying diseases, including: liver cirrhosis (80\% of cases), advanced cancer (particularly gastrointestinal and ovarian cancers, 10\% of cases), heart failure (3\%), tuberculosis (2\%), pancreatic disease (1\%), or other causes (2\%) \cite{ascites_info}. It is typically diagnosed through physical examination, medical history evaluation, and imaging techniques such as US or CT scans. 

The accumulation of fluid in the peritoneal cavity can lead to abdominal distension and discomfort, with significant implications for patient prognosis and quality of life. The management of ascites depends on the underlying cause and severity of the condition. Treatment options may include dietary modifications, medications to reduce fluid accumulation, drainage of excess fluid through paracentesis, or surgical interventions such as shunt placement or liver transplantation \cite{runyon1994care}. 

Accurate and timely assessment of ascites volume is crucial for effective disease management and monitoring response to treatment. Medical imaging techniques, including CT scans, play a vital role in visualizing and quantifying the volume of ascites, providing valuable information for clinical decision-making. Volume quantification can help determine the timing and method of treatment, assess the response to treatment, and predict disease severity.  For example, in the case of ovarian cancer, the volume of ascites at initial diagnosis of ovarian cancer is correlated with worse progression-free survival and overall survival \cite{szender2017impact}. 

Automated segmentation of ascites can be a challenging task and is hindered by its properties as a fluid. Unlike body organs, ascites cannot be inferred based on a shape prior, as it can be randomly distributed in the lower abdomen and pelvic region. Furthermore, the intensity profile of ascites is similar to that of other fluids in the abdominal region, e.g., water, urine, and bile.  

The development of automated methods for segmenting and measuring ascites volume in medical images introduces a new paradigm that is not present in the current clinical workflow. For example, accurately quantifying ascites can directly impact treatment strategies, predict overall survival, and provide essential insights for diagnosing and managing conditions such as liver cirrhosis. Moreover, the techniques refined for ascites can be applied to segmenting other fluid accumulations in the abdomen, such as hemoperitoneum, lymphocele, abscesses, bile leaks, and urinary ascites, demonstrating the potential for broader clinical use.

The aim of this study was to evaluate the effectiveness of a deep learning algorithm for the identification and volumetric quantification of ascites in targeted patient populations, specifically those with liver cirrhosis and ovarian cancer. Validation was conducted across multiple institutional datasets to evaluate model robustness. Furthermore, the model’s performance was benchmarked against expert assessments to emphasize its prospective use in clinical practice. 

\section{Materials and Methods}

\begin{figure}[!ht]
    \centering
    \includegraphics[width=0.8\linewidth]{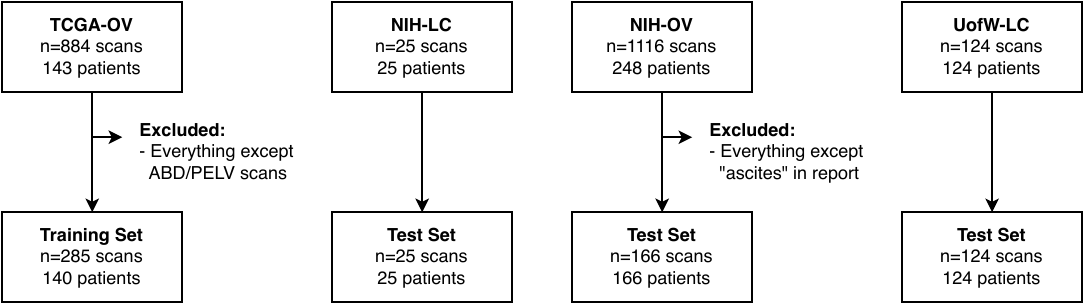}
    \caption{Flow Diagram detailing excluded scans for each dataset used in this retrospective study.}
    \label{fig:ascites-data}
\end{figure}

\begin{figure}[!ht]
     \centering
     \subfloat[]{\includegraphics[width=0.45\linewidth]
     {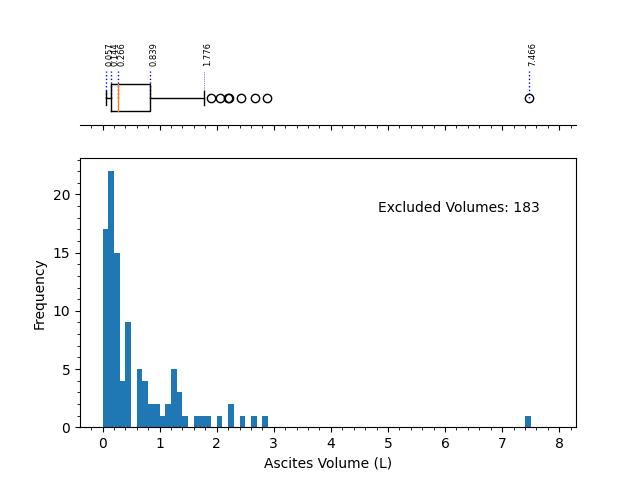}\label{<figure1>}} \hspace{4mm}
     \subfloat[]{\includegraphics[width=0.45\linewidth]
     {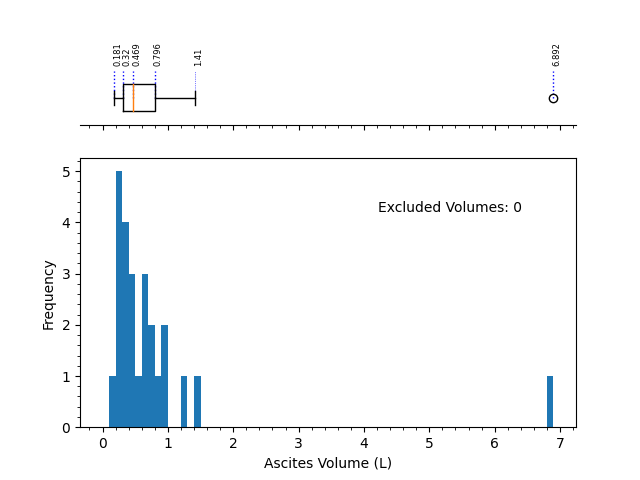}\label{<figure2>}} \\
     \subfloat[]{\includegraphics[width=0.45\linewidth]
     {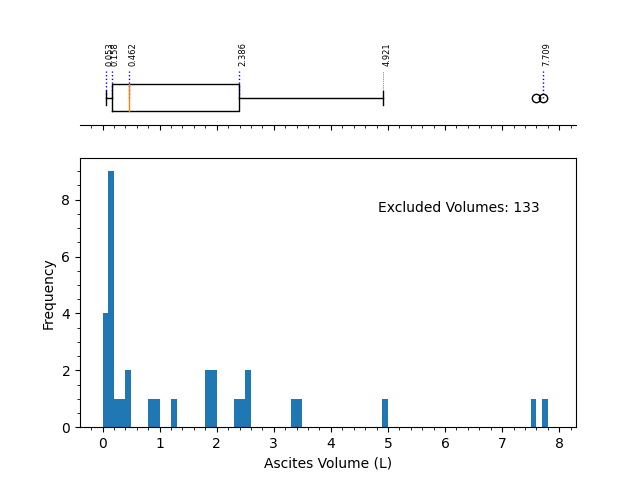}\label{<figure3>}} \hspace{4mm}
     \subfloat[]{\includegraphics[width=0.45\linewidth]
     {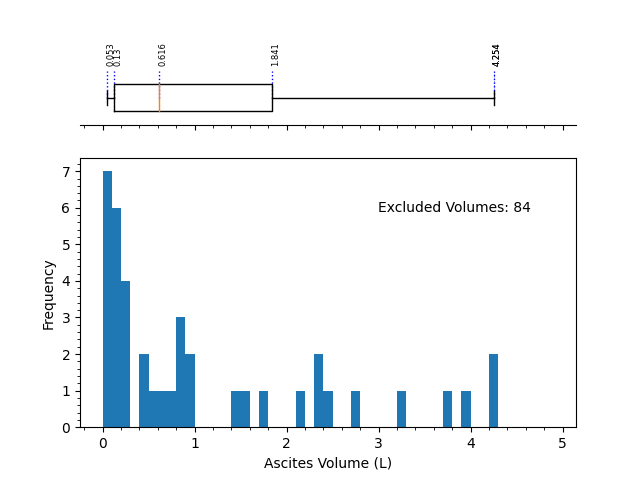}\label{<figure4>}}
     \caption{Distribution of Ascites Volume in each of the Datasets; (a) TCGA-OV, (b) NIH-LC, (c) NIH-OV, (d) UofW-LC. Note: In each dataset, volumes containing less than 50mL of detected ascites are labeled as ``trace ascites'' and designated as ``Excluded Volume''. Orange bar in each box-plot represents the median value of each dataset.}
     \label{fig:ascites_distribution}
\end{figure}

Four datasets were used in this retrospective study, which were obtained from two separate sources:  National Institutes of Health (NIH) and University of Wisconsin (UofW). Figure \ref{fig:ascites-data} provides an overview of all the datasets used, while Table \ref{tab:ascites_characteristics} and Figure \ref{fig:ascites_distribution} summarize their respective characteristics. TCGA-OV is publicly available, while the remaining three are non-public. Data from the non-public sources are Health Insurance Portability and Accountability Act-compliant. Use of the data was approved by the Institutional Review Board (IRB), and the requirement for informed consent was waived. 

\subsection{Study Datasets and Patient Characteristics}

TCGA-OV - The Cancer Genome Atlas Ovarian Cancer \cite{holback2016radiology, Clark2013}, a collaborative effort of the National Cancer Institute and the National Human Genome Research Institute, is a publicly available dataset consisting of 844 scans from 143 female participants aged between 38 and 82 years (mean: 61.0, s.d.: 11.1). This dataset plays a pivotal role in cancer research, facilitating the discovery of molecular changes and the development of prognostic biomarkers, thereby enhancing the comprehension and treatment of ovarian cancer \cite{braun2013discovery,kanchi2014integrated,islam2017hdac10}. For this study, 285 abdomen-pelvis scans of 140 patients were selected, of which 212 were intravenous contrast-enhanced, and 73 were non-contrast. Detailed information of the dataset can be found on The Cancer Imaging Archive portal page.

NIH-LC - This dataset comprised 25 scans obtained from patients with liver cirrhosis from NIH. Patients were aged between 29 and 85 years (mean: 59.4, s.d.: 13.8), with 14 males and 11 females. All scans in this dataset were intravenous contrast-enhanced. 

NIH-OV - This dataset consisted of 166 scans obtained from 166 female patients aged between 35 and 85 years (mean: 65.0, s.d.: 9.4). The scans were selected from NIH based on a clinical trial (NCT02203513) and had a mention of the keyword ``ascites'' in the radiology report. The dataset included a mixture of intravenous contrast-enhanced (n=164) and non-contrast (n=2) CT scans. 

UofW-LC - This dataset, obtained from University of Wisconsin, consisted of 124 scans and was used as an external test set from a separate institution. Scans were obtained from 124 patients (51 male, 73 female) aged between 21 and 72 years (mean: 46.0, s.d.: 12.2). The dataset included a mixture of intravenous contrast-enhanced (n=121) and non-contrast CT (n=3) scans. This study included 142 of the 406 patients whose data were previously reported \cite{doi:10.1148/ryai.210268} in a study evaluating a deep learning model for the detection of liver cirrhosis and advanced fibrosis. 

To the best of our knowledge, none of these datasets were previously used with a focus in segmenting and measuring the volume of ascites. 

\subsection{Power Analysis}

A total of 315 patients were sourced from multiple datasets for the testing cohort, of which 98 patients were identified to have ascites secondary to either ovarian cancer or liver cirrhosis. While power analysis suggests a minimum of 32 patients per condition to achieve 80\% power for detecting a medium effect size (Cohen's d=0.5) at an alpha level of 0.05, our internal liver cirrhosis test set comprised only 25 patients. However, the external liver cirrhosis test set contained 40 patients, exceeding the recommended minimum. Given that both conditions contribute to the same disease —ascites— the smaller internal test set for liver cirrhosis serves primarily as an auxiliary validation cohort. Its main purpose is to provide an initial, unbiased evaluation of the model, while the larger, external test set serves as the primary validation cohort to ensure generalizability. Considering the overall dataset size and the distribution across conditions, we believe the study remains adequately powered for its primary objectives. The flow diagram of patient selection and characteristics of the final study sample are detailed in Figure \ref{fig:ascites-data} and Table \ref{tab:ascites_characteristics}, respectively. 

\subsection{Manual Segmentation of Ascites with Active Learning}

Labeling ascites in medical images is notably challenging, primarily due to the fluid properties of ascites. However, the labeling process can be expedited through the implementation of an Active Learning schedule. Active Learning is an iterative approach within the machine learning paradigm, designed to optimize the training process by selectively choosing the most informative data samples for manual labeling. The procedure commences with a small initial batch of manually labeled data, which serves to train a preliminary model. This trained model is subsequently used to predict labels on new, unlabeled data. Any inaccuracies, such as false positives and/or negatives, are manually corrected by human experts. These newly annotated samples are then integrated into the existing training set, and the model undergoes retraining. This iterative process persists until either the desired level of performance is attained or the allocated labeling budget is exhausted. 

Each of the four datasets, whether designated for training or testing, adhered to its own Active Learning schedule, where each schedule was similar to the process detailed in Wasserthal et al. \cite{doi:10.1148/ryai.230024}. For this study, TCGA-OV was used for training, while NIH-LC, NIH-OV, and UofW-LC were used to evaluate the performance of the trained model. An initial set of 15 images was manually annotated using publicly available CT software (MitkWorkbench v2022.04). To emphasize soft tissue contrast, the intensity window was configured with a center value of 50 HU and a width of 350 HU. Manual annotations were conducted on a slice-by-slice basis using a 2D region-growing tool to ensure precise delineation of target areas. Any instances of over- or under-segmentation were corrected using paint and erase tools. After each Active Learning iteration, labels were corrected and/or refined again, using the same annotation methodology, prior to their reintegration into the active learning training pool. 

The initial manual labeling of all CT scans, as well as their subsequent corrections following each active learning cycle, were conducted by a postdoctoral fellow (B.H.). To ensure the highest level of accuracy, all training labels were verified by a board-certified radiologist with more than 10 years of clinical experience (S.L.). Furthermore, all test sets labels underwent thorough verification by a board-certified radiologist with 30 years of clinical experience in body CT image interpretation (R.M.S.). We plan to make the annotations for the training set publicly accessible. The curated annotations, and also trained segmentation model, will be published alongside this paper.   

\subsection{Detection and Automatic Volume Quantification of Ascites}

\begin{figure}[!ht]
    \centering
    \includegraphics[width=0.8\linewidth]{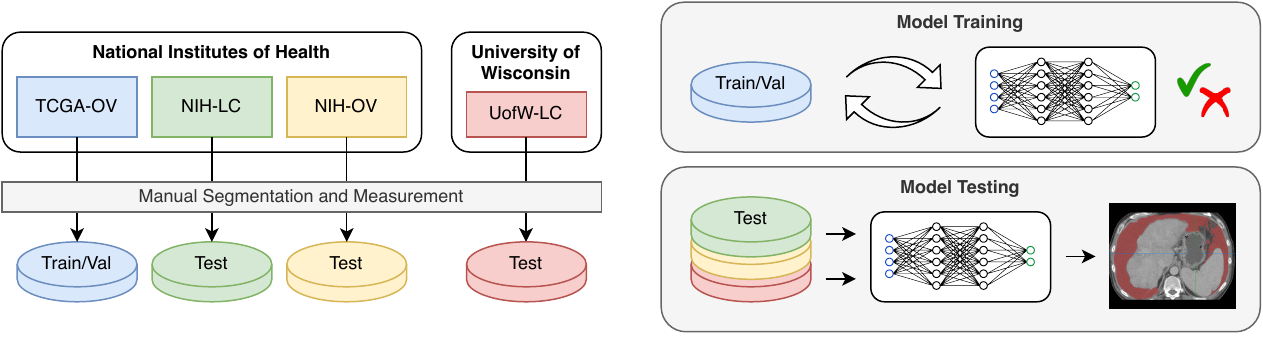}
    \caption{Experimental pipeline for automatic segmentation of ascites and volume measurement.}
    \label{fig:model}
\end{figure}

We assessed the performance of the widely recognized state-of-the-art segmentation model, nnU-Net \cite{isensee2021nnu}, for the automatic identification and volumetric quantification of ascites from CT volumes. It is important to note that nnU-Net is not an architecture per se, but rather a framework that adapts to the unique heuristics of a given dataset. It dynamically sets appropriate hyperparameters to train a 3D Residual U-Net at its core. The overview of our method and experimental pipeline is depicted in Figure \ref{fig:model}.

The segmentation model can act as an initial detector for ascites, thereby streamlining the identification of potential cases. Of note, the precise volume constituting a `trace' amount of ascites has not been universally defined. However, it is important to emphasize that the abdominal cavity, under typical physiological conditions, contains approximately 50 to 75 mL of fluid \cite{Bell2010}. Similarly, a study conducted by Yoshikawa et al., published in AJR, revealed that free fluid volumes of 10 mL or less are deemed clinically insignificant in healthy men and post-menopausal women \cite{doi:10.2214/AJR.12.9645}. This minimal accumulation is variously referred to as `free fluid', `free peritoneal fluid', or simply `trace ascites'. Notably, mild ascites in adults, which generally ranges in volume from 100 to 400 mL, may not present with overt symptoms \cite{mph2023}. 

The process of identifying ascites in this study was conducted by the model, using a predetermined threshold of 50mL. Accordingly, any volume containing less than 50mL of ascites, as determined by the segmentation mask, was categorized as "no ascites" and excluded from further volumetric analysis. This includes false positives (i.e., the model predicted ascites where the radiologist did not) and excluded false negatives (i.e., the model did not predict ascites where the radiologist did). The volume of ascites was quantified using the segmentation mask, calculated by multiplying the unit volume of the CT scan —defined by the spacing in $x\times y \times z$ dimensions— by the total number of voxels identified as ascites. Among the datasets evaluated, only NIH-OV and UofW-LC featured patients with varying ascites statuses. All patients in NIH-LC were confirmed to have ascites. 

While including false positives and excluding false negatives is not ideal, thresholding the initial predictions of the model has demonstrated high accuracy in detecting the presence of ascites. This procedure serves as an initial screening to identify outliers, with volumes labeled as ``no ascites'' marked for subsequent review by human experts. 

nnU-Net utilizes specific pre-processing steps to optimize its performance. As CT images are the modality of interest, nnU-Net applies a foreground voxel clipping process based on the 0.5 and 99.5 percentile values. This step ensures that only relevant foreground voxels are considered during the subsequent stages of the network. Additionally, since intensity values in CT images are quantitative and reflect physical properties, nnU-Net incorporates global $z$-score normalization. This normalization technique is applied uniformly to all images, allowing for standardized intensity scaling across the dataset. Finally, the loss function is a combination of binary cross-entropy and soft Dice loss with equal weighting (Equation \ref{eqn:loss}), 

\begin{equation}
    \mathcal{L}(y,\hat{y}) = 1 - \frac{2 \cdot \sum_{i=0}^{N}y_i \cdot \hat{y_i}}{\sum_{i=0}^{N}(y_i+\hat{y_i}) + \epsilon} 
    -\frac{1}{N}\sum_{i=0}^{N}\big(y_i\cdot\log{\hat{y_i}} + (1-y_i)\cdot\log(1-\hat{y_i})\big)
\label{eqn:loss}
\end{equation}

where \( y \) denotes the true binary labels and \( \hat{y} \) represents the predicted probabilities, with \( i \) indicating the index for each of the \( N \) data points. The function \( \sum \) denotes summation over all data points, and \( \epsilon \) is a small constant added to prevent division by zero in the Dice loss calculation. The Dice component of the loss, \(\frac{2 \sum_{i=0}^N y_i \hat{y}_i}{\sum_{i=0}^N (y_i + \hat{y}_i) + \epsilon}\), measures the overlap between true labels \( y \) and predicted probabilities \( \hat{y} \), adjusting for the total possible overlap plus a small constant \( \epsilon \) to avoid division by zero. The binary cross entropy component, \(\frac{1}{N} \sum_{i=0}^N \left(y_i \log(\hat{y}_i) + (1-y_i) \log(1-\hat{y}_i)\right)\), calculates the average entropy loss over all predictions, penalizing incorrect predictions based on their confidence levels. The integration of these two loss components—the Dice loss and the binary cross entropy loss—ensures that the model is effectively penalized for overlap errors as well as predictive inaccuracies related to probability distributions. This approach is particularly valuable in segmentation tasks where the object of interest is relatively small compared with the overall size of the image or volume.

By default, nnU-Net employs the stochastic gradient descent optimizer with an initial learning rate of 10$^{-2}$ and a batch size of 2. Convergence was achieved after approximately two days of training. All computational experiments were executed on an NVIDIA DGX-1 machine outfitted with 40GB A100 GPUs. The nnU-Net model generates an output in the form of a segmentation map, which serves to identify the presence of ascites within the CT volume. Furthermore, the model estimates predictive uncertainty by leveraging Softmax Confidence as a proxy measure, thereby offering insights into the confidence levels associated with the segmentation outcomes. 

% https://arxiv.org/pdf/2106.04972.pdf
% From the segmentation map, it is possible to compute the volume of ascites (in liters) as well as the average and standard deviation of the CT attenuation in Hounsfield Units (HU). 

\subsection{Model Evaluation and Statistical Analysis}

The nnU-Net model was trained and validated using all scans from TCGA-OV. Subsequently, the performance of these trained models was assessed on NIH-LC, NIH-OV and UofW-LC to evaluate their capability in segmenting ascites. Comparative metrics, including F1/Dice Score, Precision, Recall/Sensitivity, Specificity, and Volume Error, were calculated against manually annotated ascites measurements. This computation was executed using the MedPy library (v0.3.0) on Python (v3.9.10). 

Segmentation scores are presented as means with corresponding 95\% CIs, while the relative percentage error of the estimated volume is represented as a median, bounded by the 25th and 75th percentiles (IQR). This choice is informed by the data's potential skewness, outliers, or non-normal distributions (i.e. a large percentage error on a small volume is inherently less impactful compared to a large percentage error on a substantial volume).

During the training phase on TCGA-OV, a 5-fold cross-validation process was implemented, with the top-performing model from each fold being saved. This approach facilitates model ensembling and also provides a standard deviation as an uncertainty metric. The results on the test sets are reported as the mean $\pm$ standard deviation, along with 95\% confidence intervals. To evaluate discrepancies between automatic and manual measurements, multiple metrics were employed. The agreement between the automated and manual methods were quantified using the Pearson correlation coefficient ($r^2$), while Bland-Altman plots were used to assess the mean bias. Significance testing was not conducted because the comparison is across datasets rather than different methods, thus eliminating the need for such testing.

\section{Results}

\subsection{Ascites Detection}

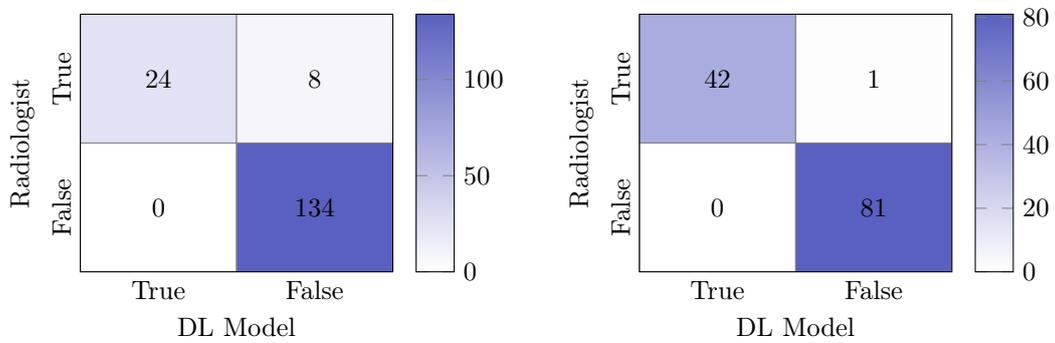
\begin{figure}[!ht]
\centering
\begin{tikzpicture}
\begin{axis}[
            scale=0.6,
            colormap={bluewhite}{color=(white) rgb255=(90,96,191)},
            xlabel= DL Model,
            ylabel=Radiologist,
            xticklabels={True, False}, % changed
            xtick={0,1}, % changed
            xtick style={draw=none},
            yticklabels={True, False,}, % changed
            ytick={0,1}, % changed
            ytick style={draw=none},
            enlargelimits=false,
            colorbar,
            yticklabel style={
              rotate=90
            },
            nodes near coords={\pgfmathprintnumber\pgfplotspointmeta},
            nodes near coords style={
                yshift=-7pt
            },
        ]
        \addplot[
            matrix plot,
            mesh/cols=2, % changed
            point meta=explicit,draw=gray
        ] table [meta=C] {
            x y C
            0 0 24
            1 0 8

            0 1 0
            1 1 134
            
        }; % added every entry where x=4 or y=4
    \end{axis}
\end{tikzpicture} 
\hspace{4mm}
\begin{tikzpicture}
\begin{axis}[
            scale=0.6,
            colormap={bluewhite}{color=(white) rgb255=(90,96,191)},
            xlabel= DL Model,
            ylabel=Radiologist,
            xticklabels={True, False}, % changed
            xtick={0,1}, % changed
            xtick style={draw=none},
            yticklabels={True, False,}, % changed
            ytick={0,1}, % changed
            ytick style={draw=none},
            enlargelimits=false,
            colorbar,
            yticklabel style={
              rotate=90
            },
            nodes near coords={\pgfmathprintnumber\pgfplotspointmeta},
            nodes near coords style={
                yshift=-7pt
            },
        ]
        \addplot[
            matrix plot,
            mesh/cols=2, % changed
            point meta=explicit,draw=gray
        ] table [meta=C] {
            x y C
            0 0 42
            1 0 1

            0 1 0
            1 1 81
            
        }; % added every entry where x=4 or y=4
    \end{axis}
\end{tikzpicture} \\ 
% (a) \hspace{7cm} (b)
\caption{Confusion matrix for deep learning based ascites detection compared with radiologist assessments on test datasets. Volume >50 mL was considered to be positive for ascites. (Left) NIH-OV (n=166), (Right) UofW-LC (n=124).}
\label{fig:ascites_detection_conf_mat}
\end{figure}

In the evaluation of ascites detection on NIH-OV and UofW-LC, the model demonstrated high accuracy. Figure \ref{fig:ascites_detection_conf_mat} and Table \ref{tab:ascites_detection_performance} presents the detection outcomes. On NIH-OV, the model achieved an accuracy of 0.952 (CI: 0.915-0.982). Although its precision was 1.000, recall was 0.750 (CI: 0.593-0.893), indicating that there were instances where ascites were not identified. This is further supported by an F1-score of 0.857 (CI: 0.744-0.944). The confusion matrix revealed 8 false negatives, which contributed to the low recall. 

On UofW-LC, the model had  an accuracy of 0.992 (CI: 0.976-1.000). The precision remained perfect at 1.000, and the recall was high at 0.977 (CI: 0.923-1.000). This resulted in a high F1-score of 0.988 (CI: 0.960-1.000). The confusion matrix for this dataset indicated one false negative finding by the model. 

\subsection{Volume Calculation}

\begin{figure}[!ht]
    \centering
    \includegraphics[width=0.25\linewidth]
    {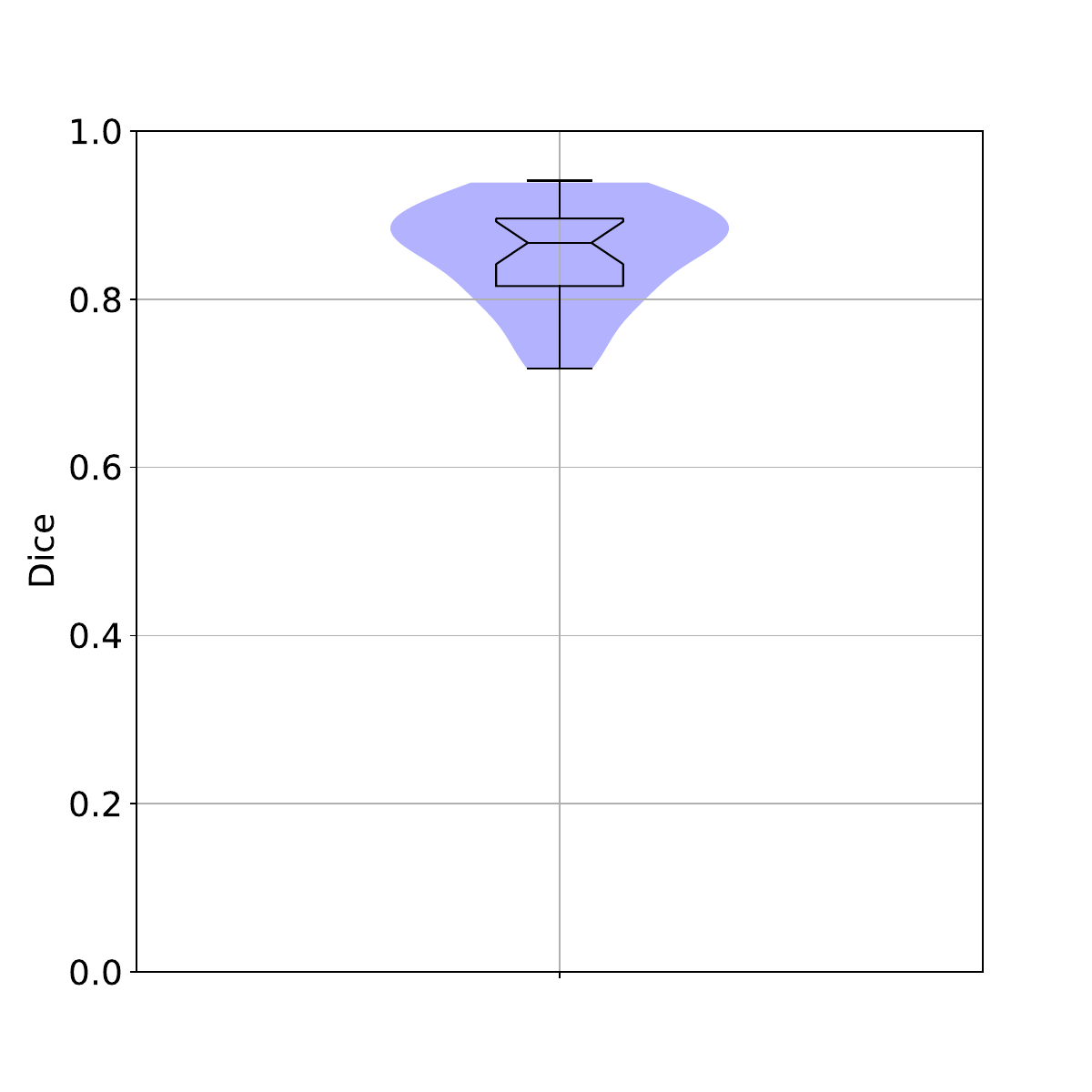} \hspace{8mm}
    \includegraphics[width=0.25\linewidth]
    {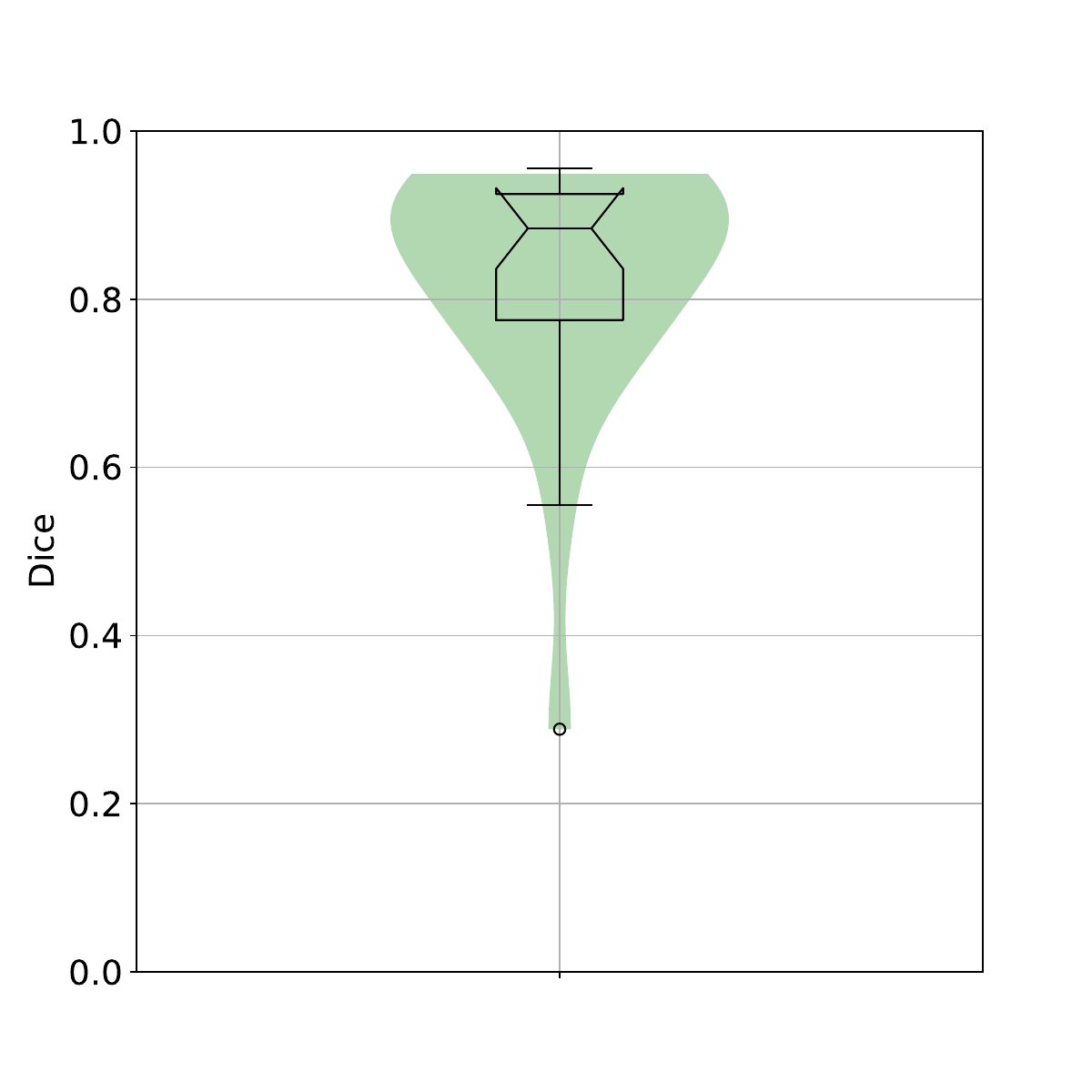} \hspace{8mm}
    \includegraphics[width=0.25\linewidth]
    {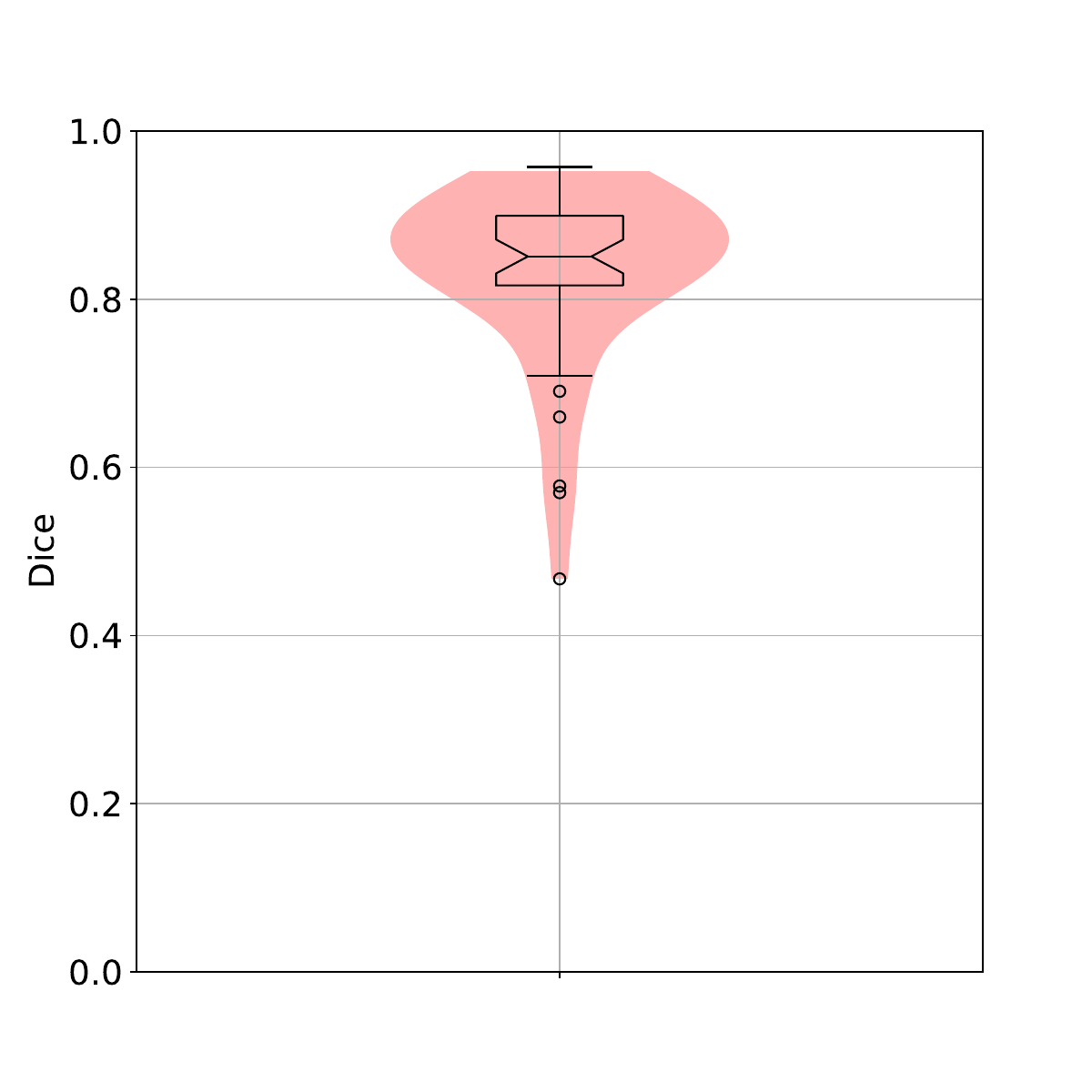} \\
    \includegraphics[width=0.3\linewidth]
    {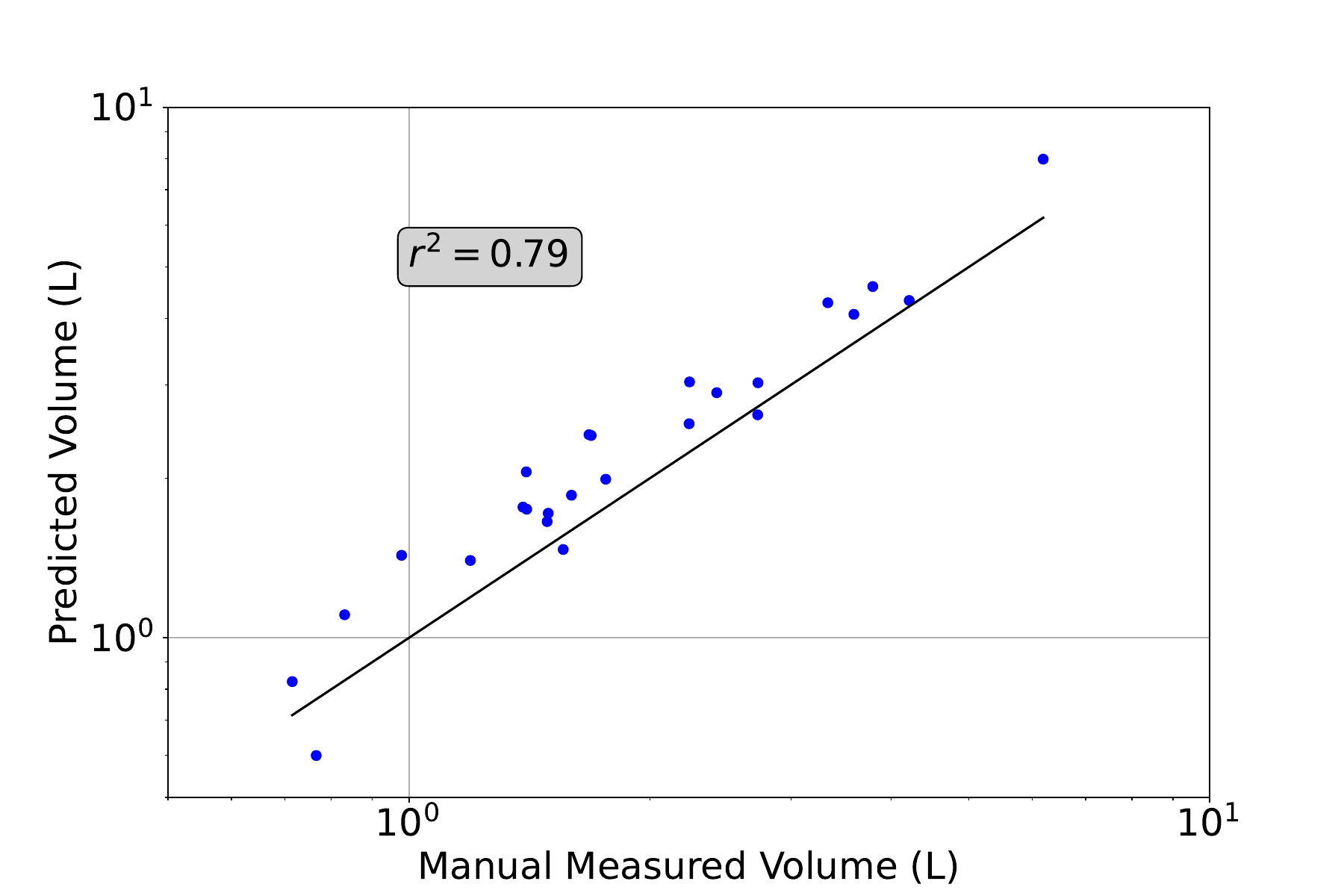} 
    \includegraphics[width=0.3\linewidth]
    {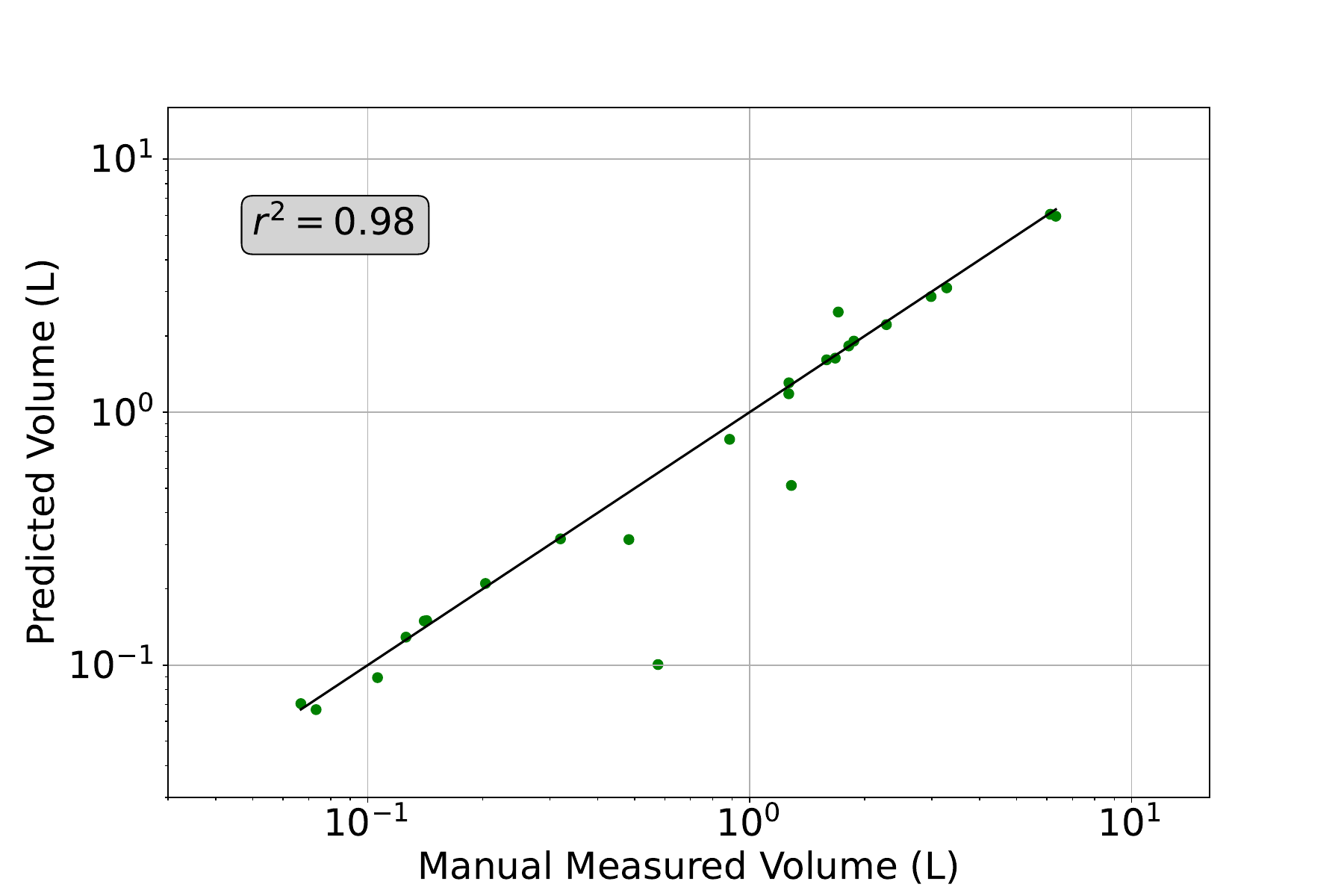} 
    \includegraphics[width=0.3\linewidth]
    {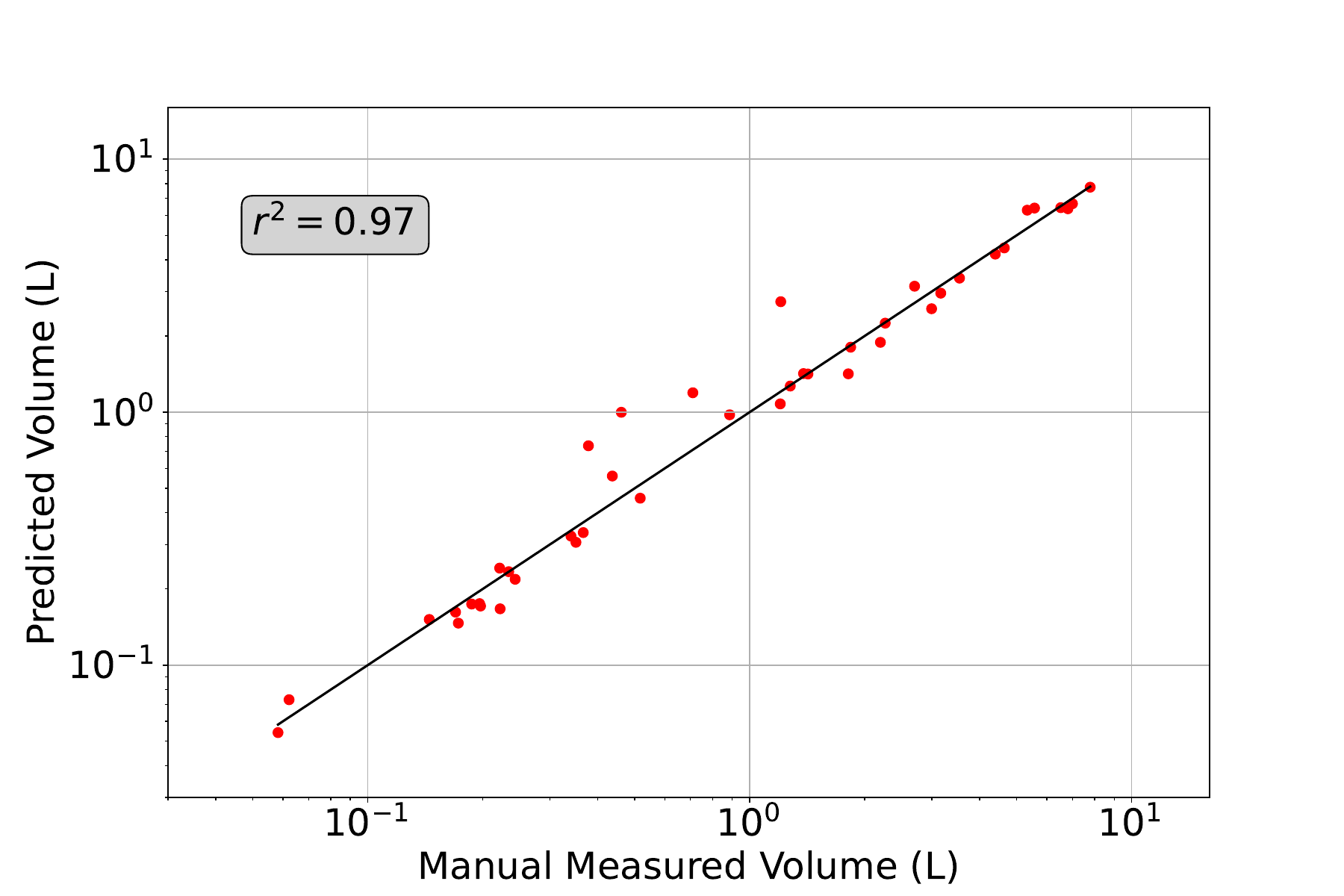} \\
    \includegraphics[width=0.3\linewidth]
    {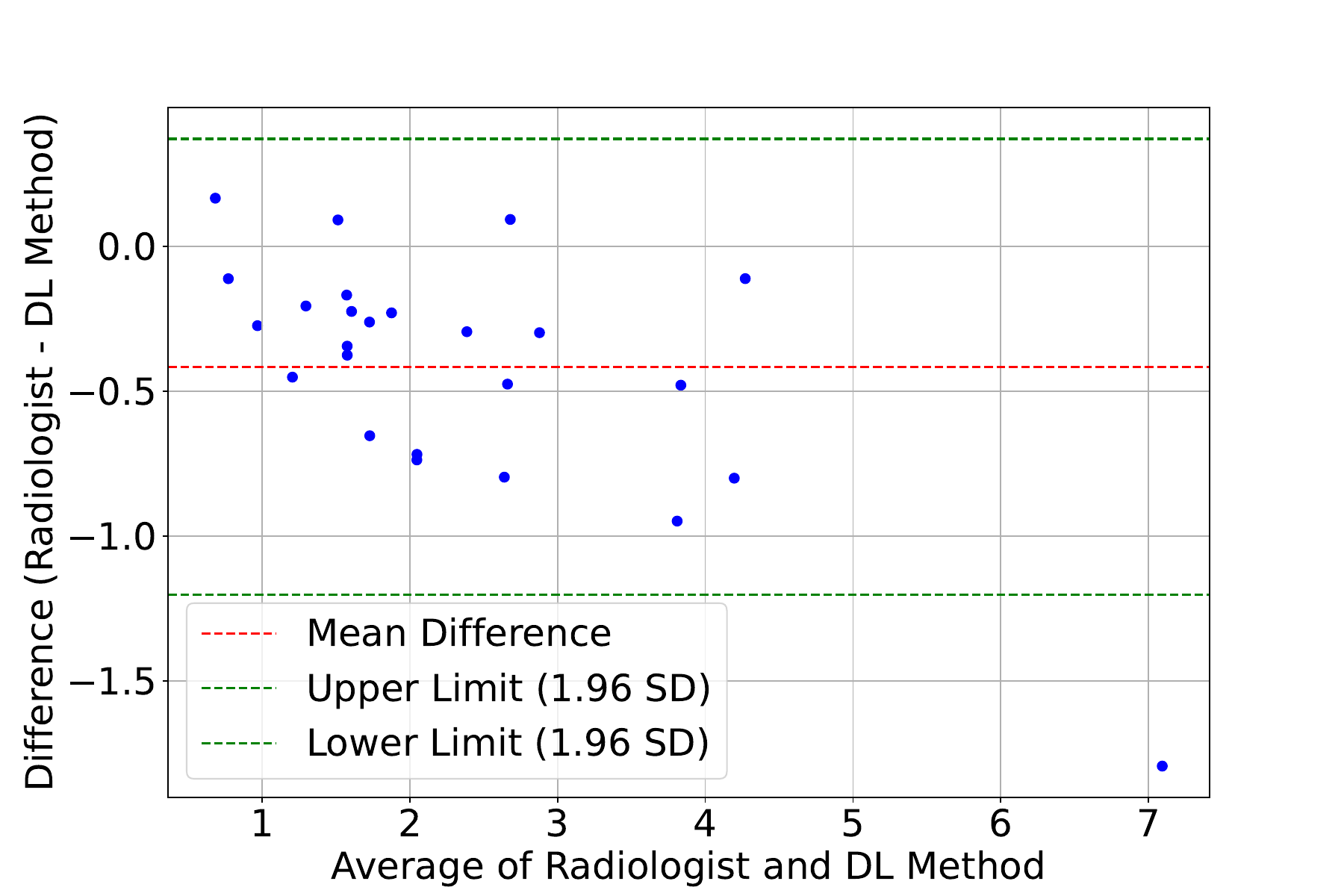}
    \includegraphics[width=0.3\linewidth]
    {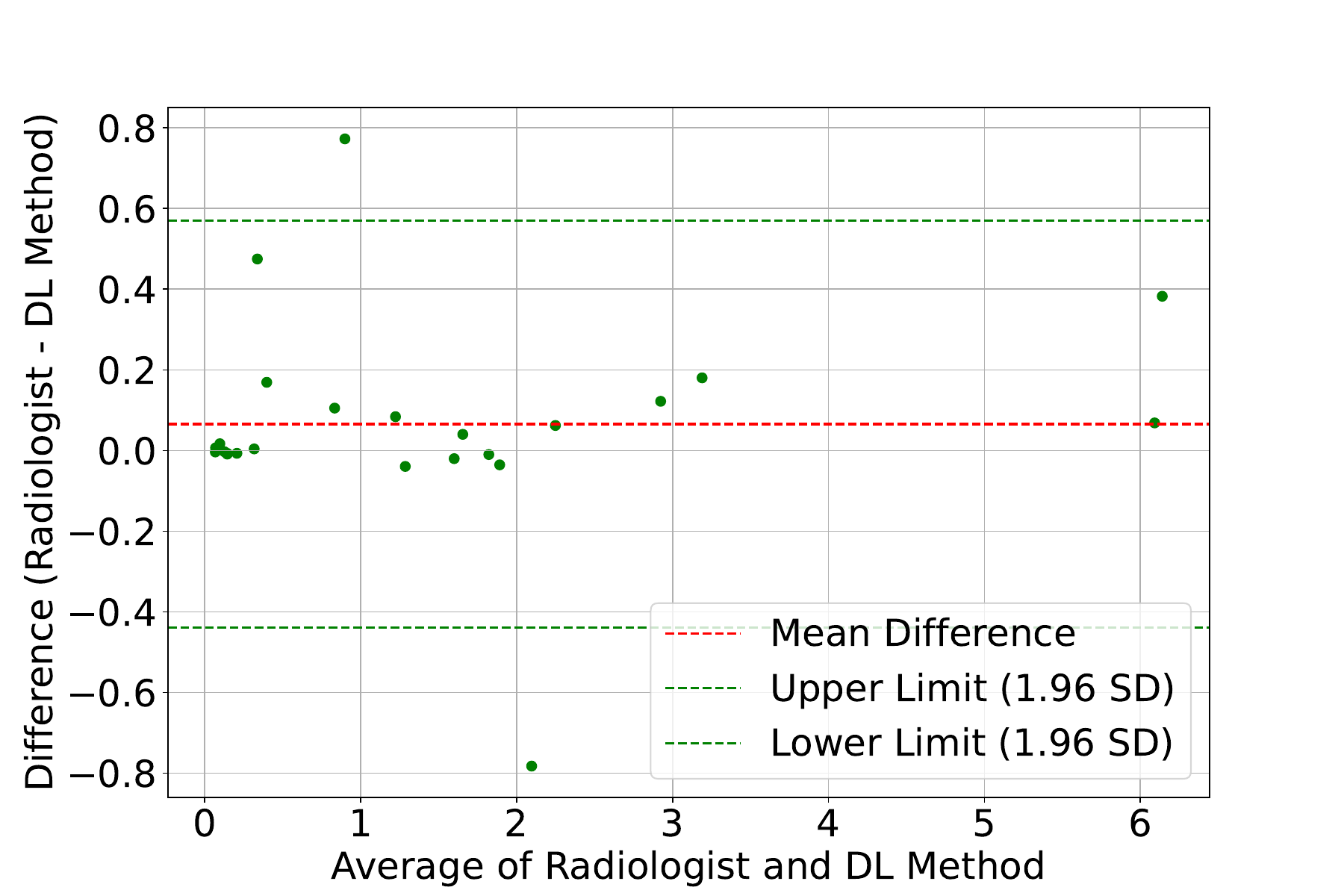}
    \includegraphics[width=0.3\linewidth]
    {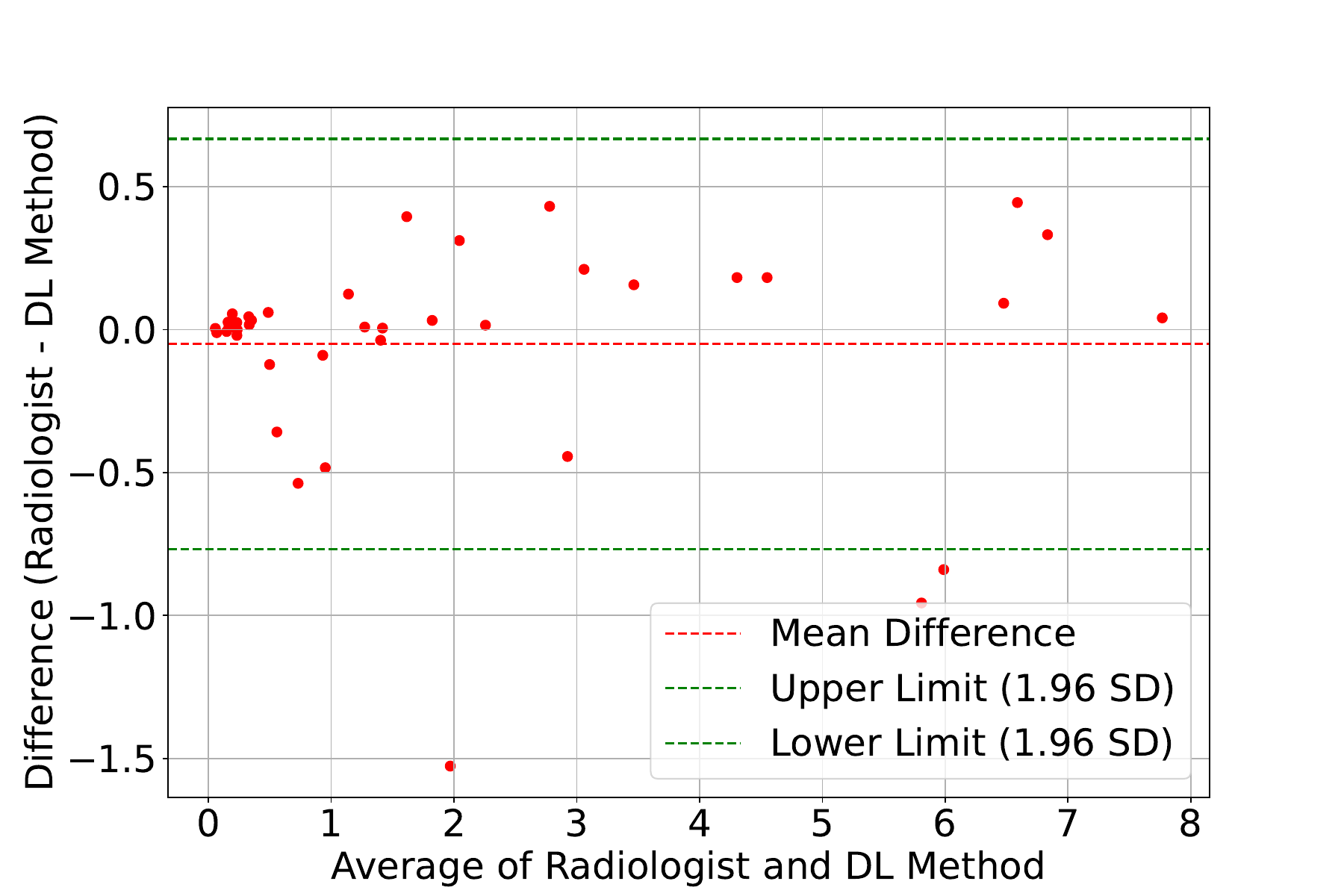} \\
    (a)\hspace{47mm}(b)\hspace{47mm}(c)
    \caption{Violin plots of Dice score distribution as well as $r^2$ and Bland-Altman plots of automatic ascites volume measurement using the deep learning (DL) model versus radiologists for each test set; (a) NIH-LC, (b) NIH-OV, (c) UofW-LC. Note: For $r^2$ plots, the axes are in log scale to account for the volume of ascites that spans several orders of magnitude.}
    \label{fig:R2_BAplot}
\end{figure}

Figure \ref{fig:R2_BAplot} and Table \ref{tab:results2} present the model's performance in volumetric analysis. For NIH-LC, the model achieved an F1/Dice score of 0.855$\pm$0.061, with a precision of 0.791$\pm$0.078, recall of 0.939$\pm$0.083, and a specificity of 0.991$\pm$0.005. The relative volume estimation error was found to be 19.6\% [IQR: 13.2, 29.0]. NIH-OV results were slightly more varied, with an F1/Dice score of 0.826$\pm$0.153, precision of 0.883$\pm$0.098, recall of 0.808$\pm$0.187, and a specificity of 0.999$\pm$0.002. The model’s volume estimation error was considerably lower at 5.3\% [IQR: 2.4, 9.7]. For UofW-LC, the model achieved an F1/Dice score of 0.830$\pm$0.107, precision of 0.832$\pm$0.150, recall of 0.847$\pm$0.085, and specificity of 0.996$\pm$0.006, accompanied by a 9.7\% [IQR: 4.5, 15.1] volume estimation error. 

The $r^2$ and Bland-Altman plots compare the deep learning (DL) model's automatic ascites volume measurements to those of a radiologist for each dataset. The model's performance, as quantified by the $r^2$ value, was 0.79 for NIH-LC. In contrast, it exhibited higher $r^2$ values of 0.98 and 0.97 for NIH-OV and UofW-LC, respectively, suggesting a strong concordance between the DL model and the radiologists' manual segmentations.

The Bland-Altman plots revealed that the model tended to over-segment the ascites volume in NIH-LC, showing a mean difference of approximately -0.5L. This tendency was largely attributed to the prevalent slice thickness of 5mm in the dataset, save for one exception with a 1mm thickness. Under such conditions, the model seemed to incur a steeper penalty. In contrast, for NIH-OV and UofW-LC, the discrepancy was primarily centered around zero, with a range spanning approximately 0.5 liters within a 1.96-standard deviation. The slice thickness in these datasets was notably more refined (predominantly 1mm thickness in NIH-OV and 3mm thickness in UofW-LC), thereby resulting in a reduced error penalty during fluid volume estimation. 

\begin{figure}[!ht]
    \centering
    \begin{minipage}{.5\textwidth}
        \centering
        \includegraphics[trim={2.5cm 2cm 2.5cm 1cm},clip,height=2.5cm]{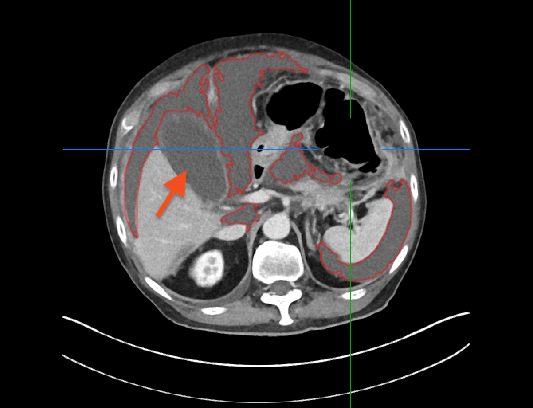} \hspace{1pt}
        \includegraphics[trim={4cm 0cm 4cm 0cm},clip,height=2.5cm]{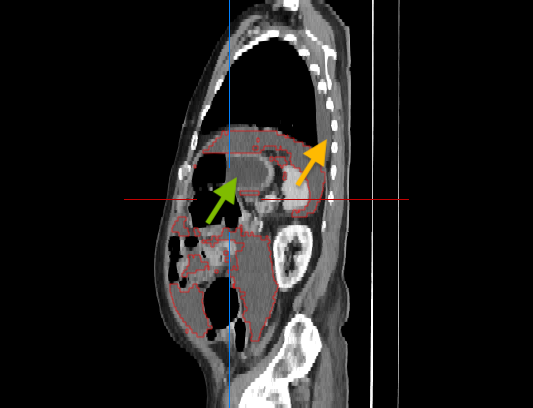} \hspace{1pt}
        \includegraphics[trim={4cm 0cm 4cm 0cm},clip,height=2.5cm]{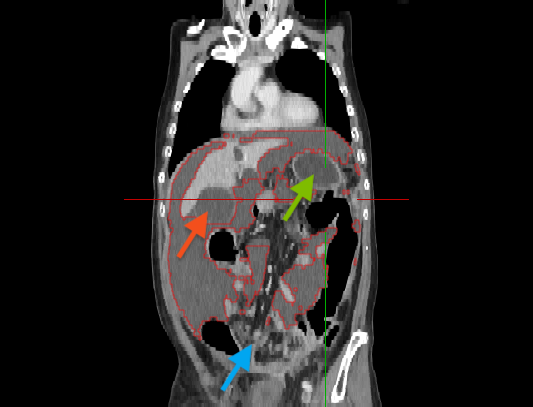} \\ 
        \includegraphics[trim={2.5cm 2cm 2.5cm 1cm},clip,height=2.5cm]{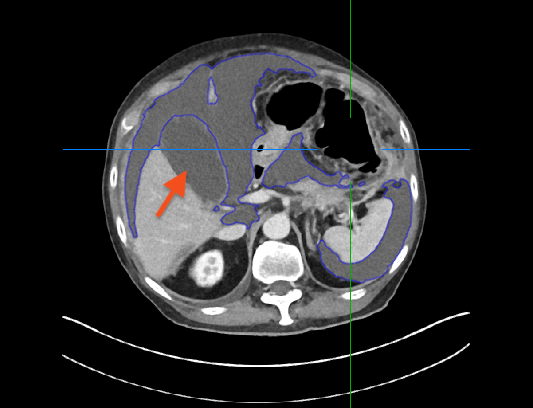} \hspace{1pt}
        \includegraphics[trim={4cm 0cm 4cm 0cm},clip,height=2.5cm]{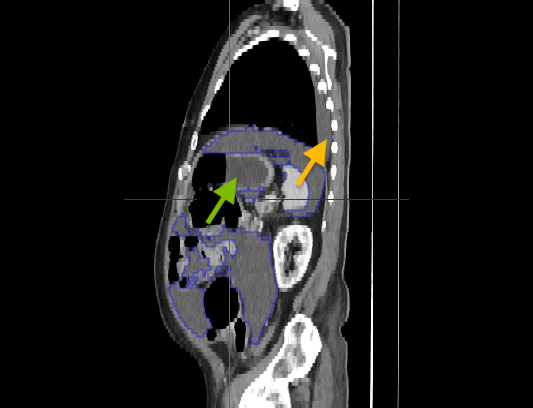} \hspace{1pt}
        \includegraphics[trim={4cm 0cm 4cm 0cm},clip,height=2.5cm]{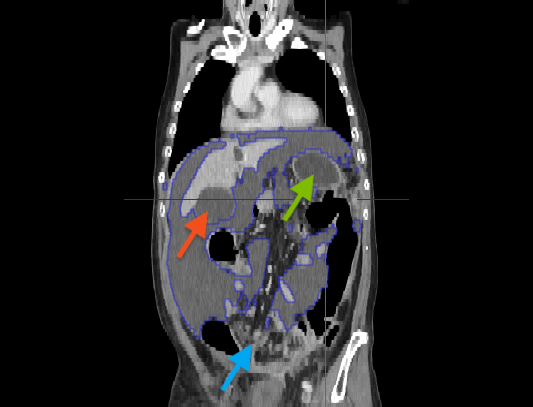} \\ 
        \textbf{(a)}
        % \caption{$dt=0.1$}
        % \label{fig:prob1_6_2}
    \end{minipage}%
    \begin{minipage}{0.4\textwidth}
        \centering
        \includegraphics[height=5cm]{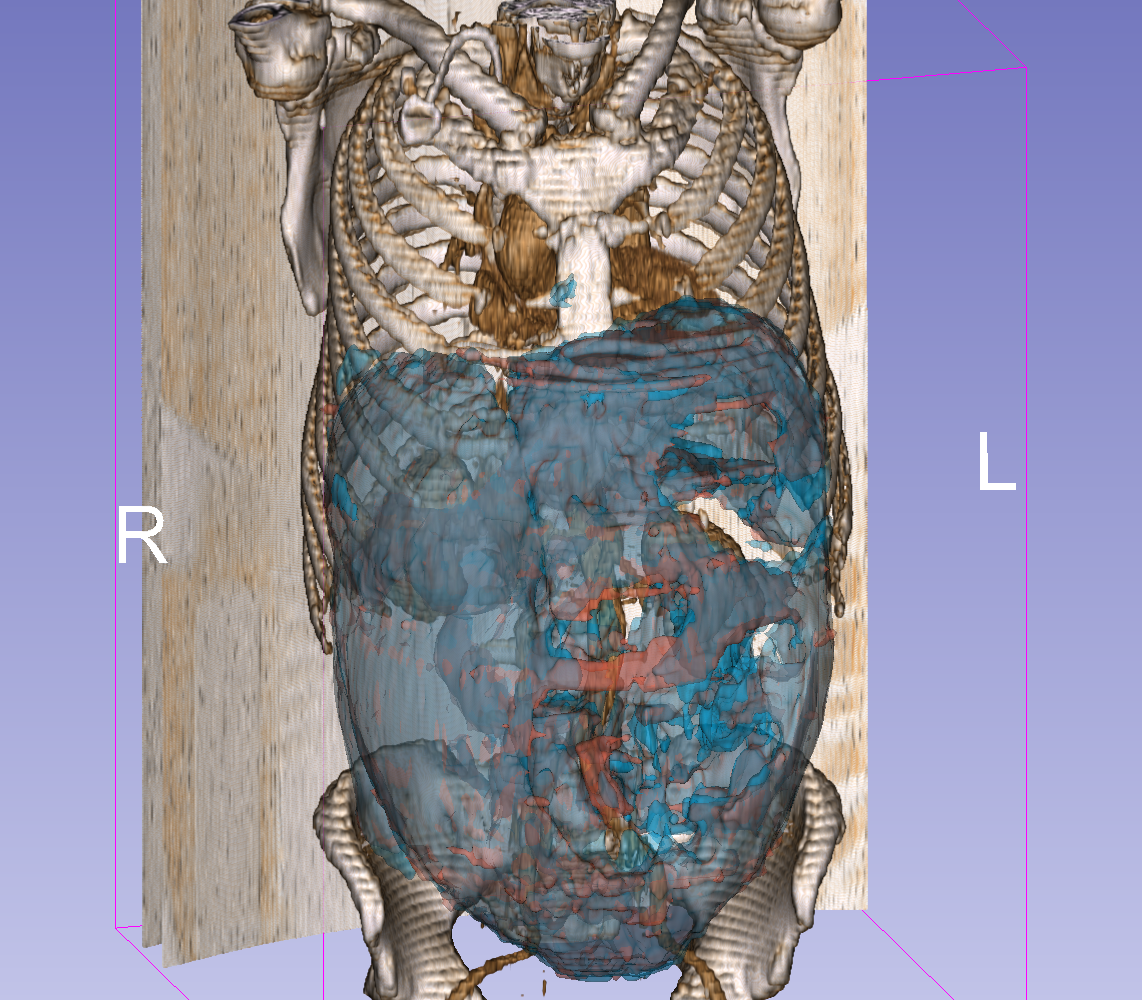} \\
        \textbf{(b)}
        % \caption{$dt =$}
        % \label{fig:case_nihlc_31}
    \end{minipage}
    \caption{Manual vs deep learning-based automatic measurement of ascites on contrast-enhanced CT images of an 85-year-old male patient with liver cirrhosis from NIH-LC. (a) Top Row: Manual annotation (red), Bottom Row: automatic annotation (blue), Left-to-Right: axial, sagittal, and coronal viewing planes. An exemplary case, showcasing the segmentation performance of the DL model on an 85-year-old male patient with liver cirrhosis. The scan was acquired with contrast, has a slice thickness of 5mm. As delineated by the arrows, the DL model selectively excluded specific regions that mimic ascites: the gallbladder (red arrow), pleural effusion (yellow arrow), and fluids within the stomach (green arrow). Additionally, the bowels were disregarded, as indicated by the blue arrow. (b) 3D rendering of ascites volume: manual measurement (red) vs automatic measurement (blue). The model achieved a segmentation Dice score of 0.920, with 13.3\% volume estimation error (4.07L actual compared with 3.59L predicted).}
    \label{fig:case_nihlc_31}
\end{figure}

\begin{figure}[!ht]
    \centering
    \includegraphics[trim={2cm 2cm 2cm 2cm},clip,width=4cm]{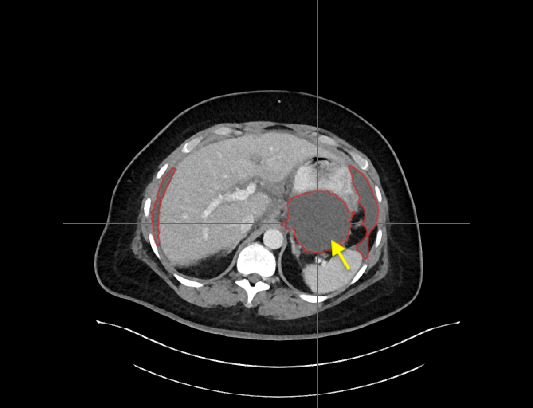} \hspace{1pt}
    \includegraphics[trim={2cm 2cm 2cm 2cm},clip,width=4cm]{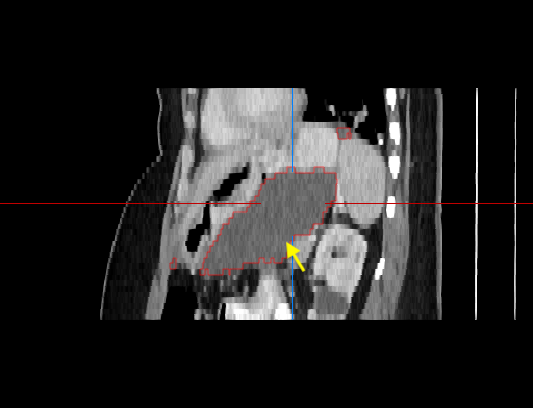} \hspace{1pt}
    \includegraphics[trim={2cm 2cm 2cm 2cm},clip,width=4cm]{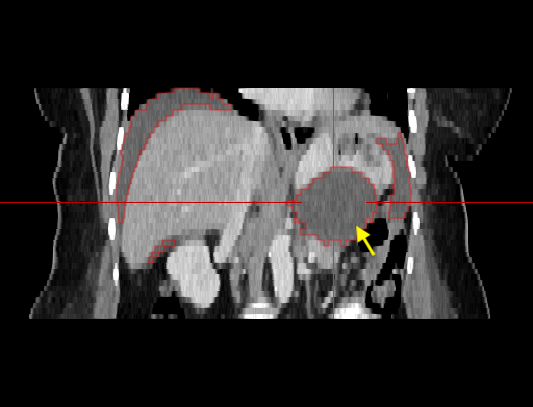} \\ 
    \includegraphics[trim={2cm 2cm 2cm 2cm},clip,width=4cm]{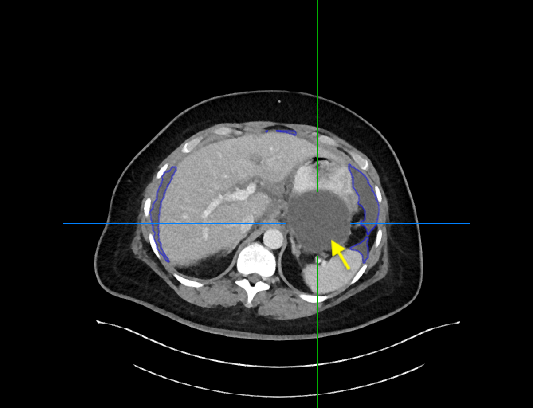} \hspace{1pt}
    \includegraphics[trim={2cm 2cm 2cm 2cm},clip,width=4cm]{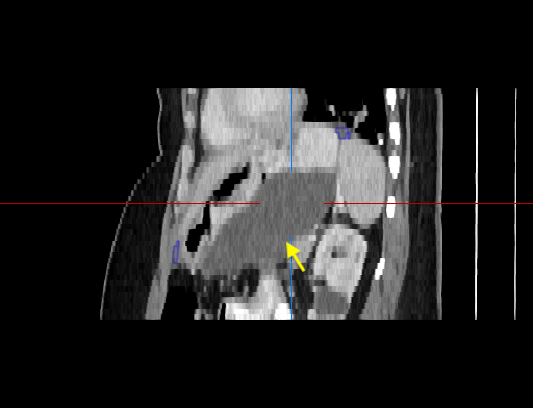} \hspace{1pt}
    \includegraphics[trim={2cm 2cm 2cm 2cm},clip,width=4cm]{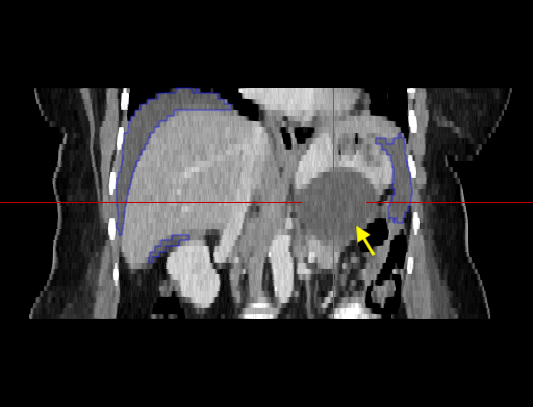} \\ 
    \caption{Manual vs deep learning-based automatic measurement of ascites on contrast-enhanced CT images of a 54-year-old female patient from NIH-LC. The images demonstrate an uncommon case of loculated ascites (yellow arrow). The scan was acquired with a 5mm slice thickness. The model achieved a segmentation Dice score of 0.717. With an estimated volume of 0.60L compared to the true volume of 0.76L, the model resulted in a 21.7\% volume estimation error. Top Row: Manual annotation (red), Bottom Row: automatic annotation (blue), Left-to-Right: axial, sagittal, and coronal viewing planes.}
    \label{fig:case_nihlc_23}
\end{figure}

\begin{figure}[!ht]
    \centering
    \includegraphics[trim={0cm 0cm 0cm 0cm},clip,width=4cm]{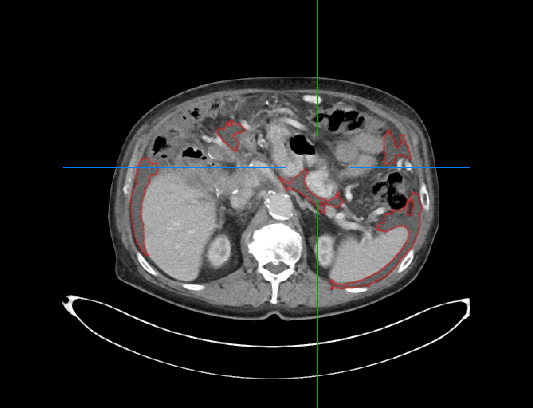} \hspace{1pt}
    \includegraphics[trim={0cm 0cm 0cm 0cm},clip,width=4cm]{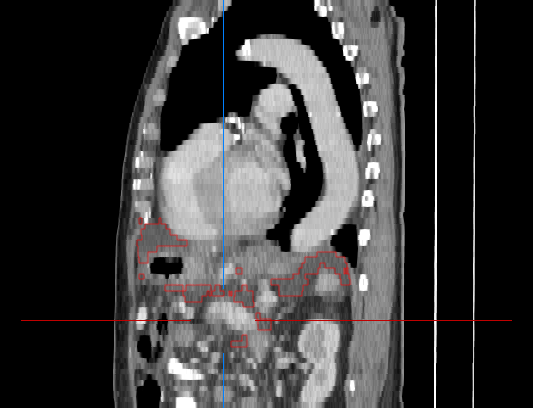} \hspace{1pt}
    \includegraphics[trim={0cm 0cm 0cm 0cm},clip,width=4cm]{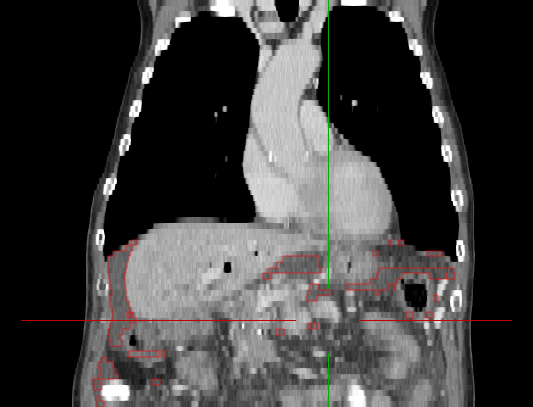} \\ 
    \includegraphics[trim={0cm 0cm 0cm 0cm},clip,width=4cm]{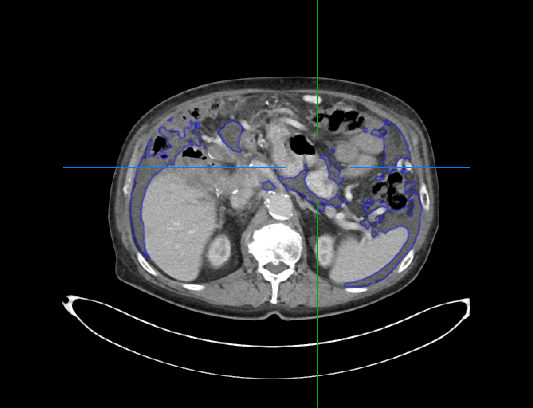} \hspace{1pt}
    \includegraphics[trim={0cm 0cm 0cm 0cm},clip,width=4cm]{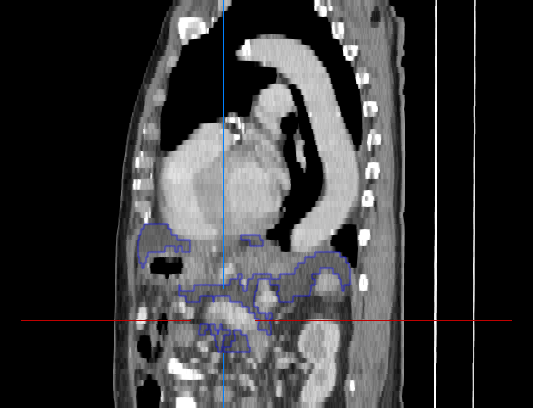} \hspace{1pt}
    \includegraphics[trim={0cm 0cm 0cm 0cm},clip,width=4cm]{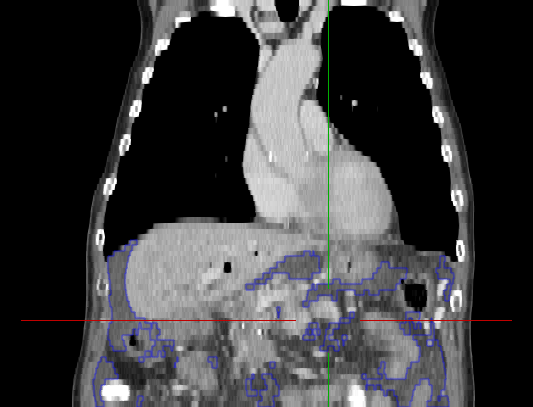} \\ 
    \caption{Manual vs deep learning-based automatic measurement of ascites on contrast-enhanced CT images of an 81-year-old male patient from NIH-LC. The images demonstrate an uncommon case mesenteric edema in front of the liver. The scan was acquired with a 5mm slice thickness. The model achieved a segmentation Dice score of 0.754. With an estimated volume of 1.43L compared to the true volume of 0.97L, the model resulted in a 46.1\% volume estimation error. Top Row: Manual annotation (red), Bottom Row: automatic annotation (blue), Left-to-Right: axial, sagittal, and coronal viewing planes.}
    \label{fig:case_nihlc_14}
\end{figure}

Figure \ref{fig:case_nihlc_31} exemplifies the deep learning model's segmentation performance in a patient with ascites, illustrating both typical and challenging cases. The model also faced difficulties with exceptional scenarios, such as patients with loculated ascites (Figure \ref{fig:case_nihlc_23}) and mesenteric edema (Figure \ref{fig:case_nihlc_14}), as observed in some cases within the NIH-LC dataset.

\section{Discussion}

In this study, we addressed the critical need for accurate and automated quantification of ascites, a condition most often resulting from either peritoneal carcinomatosis or end-stage liver disease. We employed a DL algorithm for segmentation and volumetric measurement. Notably, our results revealed a strong concordance between the algorithm’s outputs and expert radiologist evaluations. Specifically, for our internal dataset, the algorithm achieved F1-scores of 0.855 (CI: 0.831, 0.878) and 0.826 (CI: 0.764, 0.887), while for the external dataset, an F1-score of 0.830 (CI: 0.798, 0.863) was attained. These metrics substantiate the algorithm’s promising application in improving diagnostic accuracy and aiding in the clinical management of patients with ascites. 

Automated segmentation and detection of ascites do not seem to be commonly explored in literature. In 2019, Winkel et al. \cite{winkel2019evaluation} proposed a detection-based model to discriminate CT images with and without fluid. Their model achieved 85\% sensitivity and 95\% specificity in detection performance; however, additional manual analysis was required to quantify the volume for judging pathological severity. In 2022, Ko et al. \cite{ko2022deep} developed a method for automatic detection and quantification of ascites using 2D U-Net models, including U-Net, Bi-directional U-Net, Recurrent Residual U-Net, and Residual U-Net. Among these, the Residual U-Net performed the best, achieving 96\% sensitivity, 96\% specificity, and 96\% accuracy in segmentation performance. Also in 2022, Nag et al. \cite{nag2022body} developed a method for automatic detection and quantification of ascites. However, unlike Ko et al., their method was implemented in 3D to ensure segmentation continuity in adjacent CT slices. The 3D model was provided the anatomical location of each voxel, in the form of a normalized score, ranging from the chest to the lower extremities. Preliminary work by Hou et al. \cite{hou2022segmentation} demonstrated that the nnU-Net framework was capable of attaining state-of-the-art performance, establishing itself as a versatile and generalizable model for segmenting ascites, and this finding is further expanded upon in the current study. In recent literature, Wang et al. \cite{wang2023artificial} also produced similar work on ascites segmentation models. Our study sets itself apart by validating our model across multiple institutions, validating its robustness and applicability in diverse clinical settings. We anticipate that the availability of these annotations will serve as a valuable resource for researchers and practitioners aiming to further study or develop automated methods for ascites identification and quantification.   

The choice of a 50mL cutoff for ascites volume, while not a universal definition, is a pragmatic decision in our study aimed at evaluating neural network accuracy in measuring fluid volumes. Neural networks face challenges in tasks requiring high precision, notably due to limitations like data resolution and model architecture. The 50mL threshold acts as a screening tool to identify outliers, recognizing that it is one among several methods for outlier detection. 

Volumes containing less than 50mL of ascites can substantially impact the segmentation results. For instance, in cases where a patient does not have ascites but the model erroneously predicts 10mL of ascites due to the inherent inaccuracies of neural networks, the Dice score would consequently drop to 0. As shown in Table \ref{tab:ascites_characteristics}, a majority of the patients presented with less than 50mL of ascites. Such instances would significantly bias the Dice score downwards, thereby underrepresenting the capabilities of the model.

Considering the inherent challenge that ascites lacks a defined shape and can manifest in various forms, such as ``pockets'' observed in loculated ascites, missegmenting 10mL in a patient with less than 50mL of ascites disproportionately impacts the Dice score compared to a similar missegmentation in a patient with 1L of ascites. Moreover, from a clinical perspective, a patient presenting with 1L of ascites necessitates more urgent clinical attention compared with a patient with 10mL of ascites.

% \subsection{points to discuss:}
% \begin{enumerate}
%   \item[-] model does not perform as well if the patient has edema (NIH-LC-14)
%   \item[-] NIH-LC bad ones: (mesentary edema: 14 18 27 21) (large missed localized ascites: 23) (bowel: 29) (questionable edema: 19)
% \end{enumerate}

The model also faces challenges In exceptional medical cases, one being loculated ascites. In a typical case of ascites, fluid accumulates freely in the peritoneal cavity. However, in loculated ascites, the fluid is confined to specific compartments within the abdomen due to adhesions, inflammation, or scarring. Despite loculated ascites being a form of ascites and medically defined as such, it is generally less common than free-flowing ascites. Its incidence may be higher in specific populations, such as those with peritoneal carcinomatosis, post-surgical adhesions, or peritonitis. In our study, it was observed that the model struggled to perform well on these uncommon cases. 

Another special circumstance involves the presence of mesenteric edema, which, like ascites, exhibits similar radiologic characteristics, complicating their differentiation. On various imaging modalities, both conditions can manifest as areas with increased fluid density. The similar appearance of fluid accumulation in both scenarios poses challenges in distinguishing one from the other. Further complexity arises when ascites is localized within the mesentery, effectively mimicking the appearance of mesenteric edema. Moreover, there are instances where both conditions coexist, thereby further obscuring the distinction between them and complicating the task for even experienced radiologists to accurately differentiate solely based on imaging. Other special circumstances, such as hepatic hydrothorax, hemoperitoneum, foreign bodies, trauma, etc., cannot be confirmed with the datasets used in this study since the necessary indicators are absent. The scope of specific corner cases to test is vast, and many lack the clinical data —or sufficient clinical data— to achieve statistical significance. However, it opens potential for future work.

The task of accurately segmenting ascites presents a substantial challenge for radiologists, not only due to the inherent difficulty of the task but also because of the extensive time commitment required to segment ascites accurately for even a single patient. Despite these challenges, the precise quantification of ascites volume holds potential for various clinical applications and future research avenues. For instance, the extent of ascites can be used to classify liver function according to the Child Pugh Score, which is instrumental in predicting patient outcomes such as overall survival.

This study, while informative, has several limitations that merit discussion. A consideration for enhancing this study in future iterations is the incorporation of annotations from multiple expert radiologists for reference standard labeling. The use of labels verified by multiple experts would enhance the robustness of the findings, but securing the participation of multiple experts for label verification presents substantial resource challenges. Another limitation pertains to model generalizability. The dataset primarily used for training the model (TCGA-OV) was sourced from an ovarian cancer cohort, which is predominantly female and focused on a single disease type. This raises questions about the model’s applicability across diverse patient populations and other types of ascites. Therefore, follow-up work should aim to train the model on more diverse datasets, encompassing different sexes and underlying conditions, to improve its generalizability. 

In summary, our study underscores the efficacy of DL algorithms in the precise identification and volumetric quantification of ascites, showing substantial agreement with assessments by expert radiologists. However, the study does reveal specific limitations that warrant attention in future research. Addressing these issues would not only bolster the model’s robustness but also extend its applicability in real-world clinical scenarios, thus facilitating more effective diagnosis and management of conditions related to ascites. 

\newpage

\bibliographystyle{vancouver}
\bibliography{references}

\newpage

\section{Tables}

% Please add the following required packages to your document preamble:
% \usepackage{booktabs}
\begin{table}[H]
\centering
\caption{Characteristics of the Training and Test Datasets. Note: For ascites volume, statistics were calculated omitting values less than 50mL. Both institutions used CT scanners from GE Medical Systems (LightSpeed and subsequent models), Canon (Acquilion) and Siemens (Somatom, Definition, and subsequent models). `N' represents the number of scans, while `P' denotes the number of patients. Statistics for age and sex are stratified by patient. F = female, M = male.}
\begin{tabular}{@{}lcccc@{}}
\toprule
                        & ~ TCGA-OV      ~    & ~ NIH-LC   ~      & ~ NIH-OV   ~       & ~ UofW-LC  ~       \\
                        & ~ Training     ~    & ~ Testing  ~      & ~ Testing  ~       & ~ Testing  ~       \\
                        & ~ (n=285)      ~    & ~ (n=25)   ~      & ~ (n=166)  ~       & ~ (n=124)  ~       \\ 
                        & ~ (p=140)      ~    & ~ (p=25)   ~      & ~ (p=166)  ~       & ~ (p=124)  ~       \\ 
\midrule    
~ Age                   &                     &                   &                    &                    \\
~ \quad mean + s.d.     & 60.3$\pm$11.0       & 59.4$\pm$13.8     & 65.0$\pm$9.4       & 64.0$\pm$12.2      \\
~ \quad median          & 60                  & 62                & 67                 & 47                 \\
~ \quad IQR             & [51, 68]            & [52, 68]          & [59, 70]           & [36, 54]           \\
~ \quad range           & [38, 81]            & [29, 85]          & [35, 85]           & [21, 72]           \\
\midrule    
~ Sex                   &                     &                   &                    &                    \\
~ \quad M               & 0                   & 14                & 0                  & 51                 \\
~ \quad F               & 140                 & 11                & 166                & 73                 \\
\midrule    
~ Contrast              &                     &                   &                    &                    \\
~ \quad Yes             & 212                 & 25                & 164                & 121                \\
~ \quad No              & 73                  & 0                 & 2                  & 3                  \\
\midrule    
~ Slice Thickness (mm)  &                     &                   &                    &                    \\
~ \quad mean + s.d.     & 5.223$\pm$1.189     & 4.840$\pm$0.800   & 1.024$\pm$0.310    & 3.581$\pm$1.123    \\
~ \quad median          & 5.0                 & 5.0               & 1.0                & 3.0                \\
~ \quad IQR             & [5.0, 5.0]          & [5.0, 5.0]        & [1.0, 1.0]         & [3.0, 5.0]         \\
~ \quad range           & [0.8, 10.0]         & [1.0, 5.0]        & [1.0, 5.0]         & [2.0, 10.0]        \\
\midrule    
~ Pixel Spacing (mm)    &                     &                   &                    &                    \\
~ \quad mean + s.d.     & 0.750$\pm$0.088	  & 0.844$\pm$0.089   & 0.820$\pm$0.070    & 0.788$\pm$0.093    \\
~ \quad median          & 0.742               & 0.858             & 0.816              & 0.781              \\
~ \quad IQR             & [0.703, 0.781]      & [0.782, 0.920]    & [0.781, 0.850]     & [0.703, 0.849]     \\
~ \quad range           & [1.030, 0.527]      & [0.691, 0.977]    & [0.664, 0.977]     & [0.625, 0.977]     \\
\midrule    
~ No. Slices            &                     &                   &                    &                    \\
~ \quad mean + s.d.     & 95.0$\pm$46.4       & 81.0$\pm$75.3     & 626$\pm$62         & 111$\pm$45         \\
~ \quad median          & 89                  & 67                & 631                & 100                \\
~ \quad IQR             & [79, 98]            & [46, 81]          & [602, 655]         & [85, 156]          \\
~ \quad range           & [20, 597]           & [39, 426]         & [126, 707]         & [35, 228]          \\
\midrule    
~ Ascites Vol.          &                     &                   &                    &                    \\
~ \quad n<50ml          & 183                 & 0                 & 133                & 84                 \\
~ \quad\quad mean + s.d.& 0.644$\pm$0.936     & 0.831$\pm$1.303   & 1.539$\pm$1.983    & 1.148$\pm$1.290    \\
~ \quad\quad median     & 0.266               & 0.469             & 0.462              & 0.616              \\
~ \quad\quad IQR        & [0.144, 0.839]      & [0.320, 0.796]    & [0.158, 2.386]     & [0.103, 1.841]     \\ 
~ \quad\quad range      & [0.057, 7.466]      & [0.181, 6.892]    & [0.053, 7.709]     & [0.053, 4.254]     \\ 
\bottomrule
\end{tabular}
\label{tab:ascites_characteristics}
\end{table}

\newpage

% Please add the following required packages to your document preamble:
% \usepackage{booktabs}
\begin{table}[H]
\centering
\caption{Detection performance for deep learning-based ascites detection compared with radiologist assessments on test datasets. Note: Data in parentheses are 95\% Confidence Intervals obtained via bootstrap sampling (number of samples = 10,000).}
\begin{tabular}{@{}lcccc@{}}
\toprule
~~          & Accuracy              & Precision             & Recall                & F1                    ~~\\ 
\midrule
~~NIH-OV    & 0.952 (0.916, 0.982)  & 1.000 (1.000, 1.000)  & 0.750 (0.593, 0.893)  & 0.857 (0.744, 0.944)  ~~\\
~~UofW-LC   & 0.992 (0.976, 1.000)  & 1.000 (1.000, 1.000)  & 0.977 (0.923, 1.000)  & 0.988 (0.960, 1.000)  ~~\\ 
\bottomrule
\end{tabular}
\label{tab:ascites_detection_performance}
\end{table}

% bootstrap_ci with 10000 samples
% NIH-OV (brackets denote 95% confidence intervals)
% accuracy:     0.9518072289156626 (0.9156626506024096, 0.9819277108433735)
% precision:    1.0 (1.0, 1.0)
% recall:       0.75 (0.5925925925925926, 0.8928571428571429)
% f1:           0.8571428571428571 (0.7441860465116279, 0.9444696969696966)
% tn, fp, fn, tp
% (134, 0, 8, 24)
%
% UofW (brackets denote 95% confidence intervals)
% accuracy:     0.9919354838709677 (0.9758064516129032, 1.0),
% precision:    1.0 (1.0, 1.0),
% recall:       0.9767441860465116 (0.9230769230769231, 1.0),
% f1:           0.988235294117647 (0.9600000000000001, 1.0))
% tn, fp, fn, tp
% (81, 0, 1, 42)

% Please add the following required packages to your document preamble:
% \usepackage{booktabs}
\begin{table}[H]
\centering
\caption{Quantitative results of the ascites segmentation model on test datasets. Note: F1/Dice, Precision, Recall/Sensitivity (Rec./Sen.), and Specificity are given as means $\pm$ SDs, with 95\% Confidence Intervals in brackets. The percentage error measures the discrepancy between the volume estimated by the deep learning model and the true volume, expressed as a percentage of the true volume. Given the substantial variability in ascites volume, results are displayed as medians, accompanied by the 25th and 75th percentiles in brackets.}
% \resizebox{\textwidth}{!}{
\begin{tabular}{@{}l|cccc|c@{}}
\toprule
                & F1/Dice          & Precision        & Rec./Sen.        & Specificity      & \% Error     ~~\\ \midrule
~~NIH-LC        & 0.855$\pm$0.061  & 0.791$\pm$0.078  & 0.939$\pm$0.083  & 0.991$\pm$0.005  & 19.6         ~~\\
~~              & [0.831, 0.878]   & [0.905, 0.971]   & [0.760, 0.821]   & [0.997, 0.999]   & [13.2, 29.0] ~~\\ 
~~NIH-OV        & 0.826$\pm$0.153  & 0.883$\pm$0.098  & 0.808$\pm$0.187  & 0.999$\pm$0.002  & 5.3          ~~\\
~~              & [0.764, 0.887]   & [0.844, 0.923]   & [0.733, 0.883]   & [0.998, 1.000]   & [2.4, 9.7]   ~~\\ 
~~UofW-LC       & 0.830$\pm$0.107  & 0.832$\pm$0.150  & 0.847$\pm$0.085  & 0.996$\pm$0.006  & 9.7          ~~\\ 
~~              & [0.798, 0.863]   & [0.787, 0.878]   & [0.822, 0.873]   & [0.994, 0.998]   & [4.5, 15.1]  ~~\\ \bottomrule
\end{tabular}
% }
\label{tab:results2}
\end{table}

\newpage

\section{Figures}

\clearpage

\begin{figure}[!ht]
    \centering
    \includegraphics[trim={0cm 0cm 0cm 0cm},clip,width=4cm]{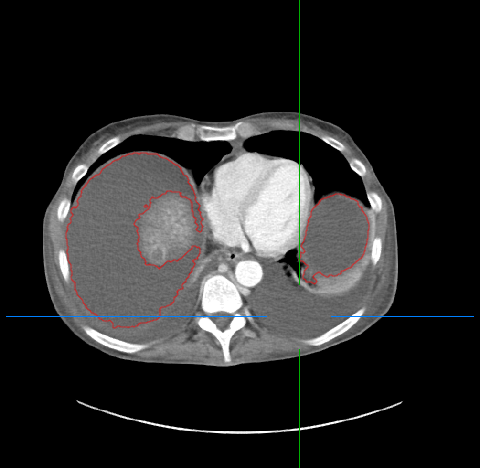} \hspace{1pt}
    \includegraphics[trim={0cm 0cm 0cm 0cm},clip,width=4cm]{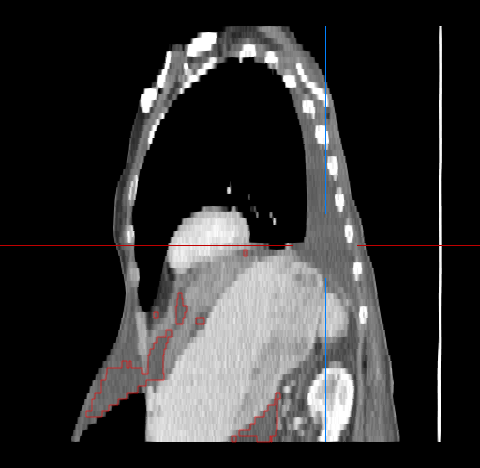} \hspace{1pt}
    \includegraphics[trim={0cm 0cm 0cm 0cm},clip,width=4cm]{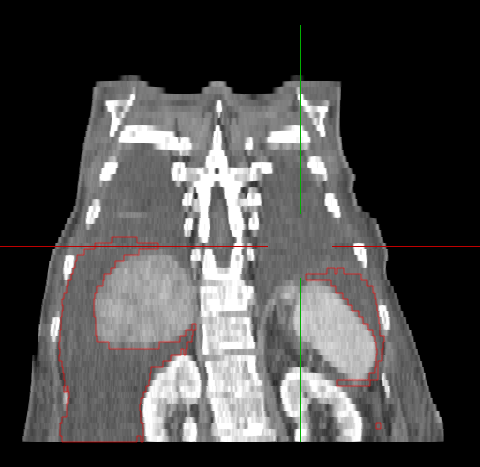} \\ 
    \includegraphics[trim={0cm 0cm 0cm 0cm},clip,width=4cm]{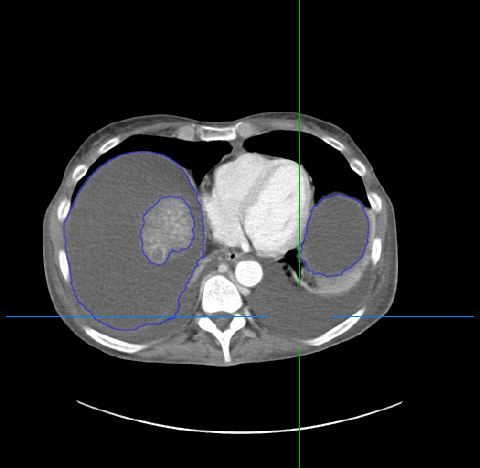} \hspace{1pt}
    \includegraphics[trim={0cm 0cm 0cm 0cm},clip,width=4cm]{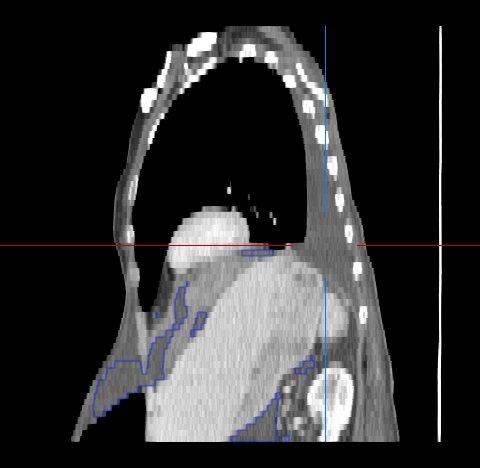} \hspace{1pt}
    \includegraphics[trim={0cm 0cm 0cm 0cm},clip,width=4cm]{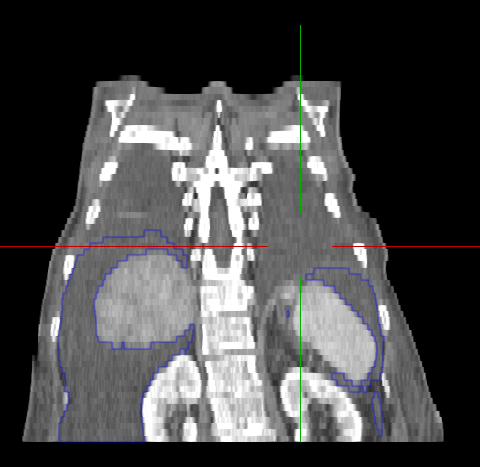} \\ 
    \caption{Manual vs automatic measurement of ascites on scan from NIH-LC by DL model, 68-year-old female imaged with contrast. The model is capable of distinguishing ascites from pleural effusion. Top Row: Manual annotation (red), Bottom Row: automatic annotation (blue), Left-to-Right: axial, sagittal, and coronal viewing planes.}
    \label{fig:case_nihlc_20}
\end{figure}

\begin{figure}[!ht]
    \centering
    \includegraphics[trim={0cm 2cm 0cm 2cm},clip,width=4cm]{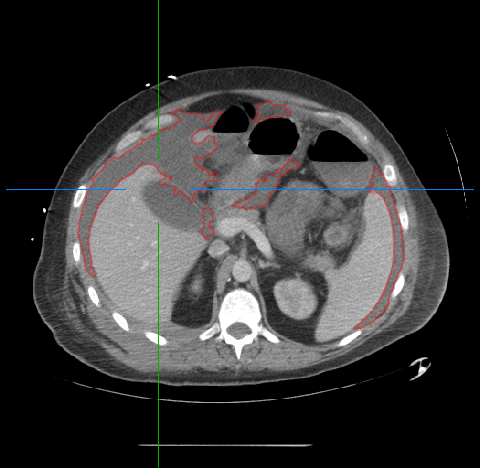} \hspace{1pt}
    \includegraphics[trim={0cm 2cm 0cm 2cm},clip,width=4cm]{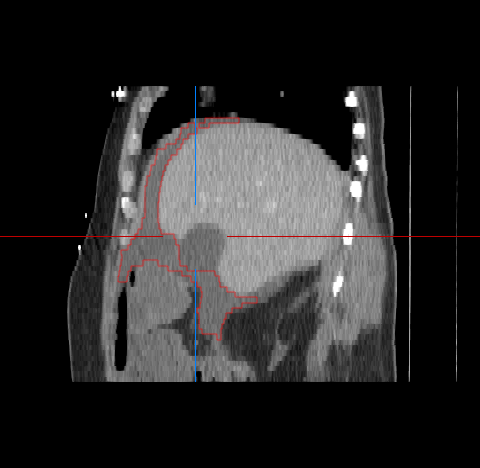} \hspace{1pt}
    \includegraphics[trim={0cm 2cm 0cm 2cm},clip,width=4cm]{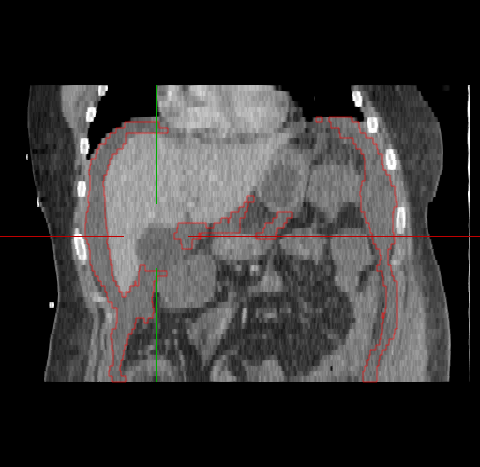} \\ 
    \includegraphics[trim={0cm 2cm 0cm 2cm},clip,width=4cm]{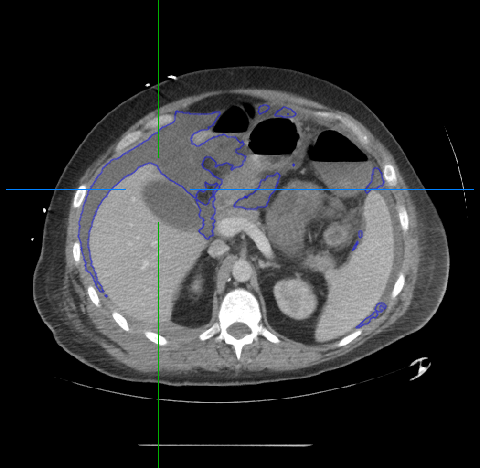} \hspace{1pt}
    \includegraphics[trim={0cm 2cm 0cm 2cm},clip,width=4cm]{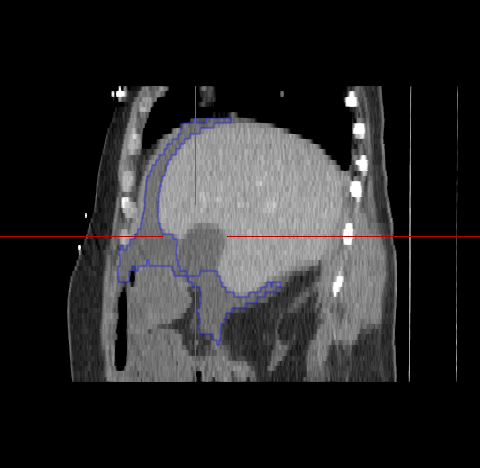} \hspace{1pt}
    \includegraphics[trim={0cm 2cm 0cm 2cm},clip,width=4cm]{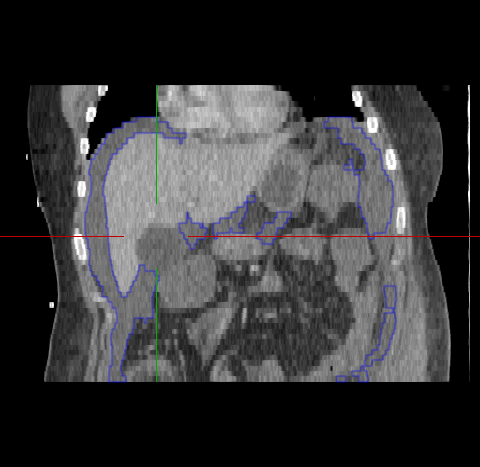} \\ 
    \caption{Manual vs automatic measurement of ascites on scan from NIH-LC by DL model, 36-year-old male imaged with contrast. The model is capable of distinguishing ascites from gallbladder. Top Row: Manual annotation (red), Bottom Row: automatic annotation (blue), Left-to-Right: axial, sagittal, and coronal viewing planes.}
    \label{fig:case_nihlc_29}
\end{figure}

\end{document}